\documentclass[a4paper,twocolumn]{igs}

\usepackage{natbib}
\usepackage{stfloats}
\usepackage{amsmath}
\usepackage{amssymb}
\usepackage{textcomp}
\usepackage{accents}
\usepackage{bm}
\usepackage{graphicx}
\usepackage{color}
\usepackage{mathtools}

\makeatletter
\newcommand{\Tr}{\mathop{\operator@font Tr}\nolimits}
\makeatother

\newcommand{\eg}{\textit{e.g.}}
\newcommand{\ie}{\textit{i.e.}}
\newcommand{\dd}{\mathrm{d}}
\newcommand{\Dd}{\mathrm{D}}
\newcommand{\ee}{\mathrm{e}}
\newcommand{\ubb}{\bar{\bm{u}}}
\newcommand{\rhob}{\bar{\rho}}
\newcommand{\LF}{_{\Lambda\rightarrow\Phi}}
\newcommand{\LP}{_{\Lambda\rightarrow\Pi}}
\newcommand{\FP}{_{\Phi\rightarrow\Pi}}
\newcommand{\PF}{_{\Pi\rightarrow\Phi}}

\newcommand{\nablab}{\bm{\nabla}}

\newcommand{\bss}{\bm{S}_b}
\newcommand{\ms}{\ensuremath{\mathrm{m\,s}^{-1}}}

\begin{document}
		
\title[Comments on RKE-based avalanche models]
      {Comments on avalanche flow models based on the concept of random
       kinetic energy}

\author[Issler and others]{Dieter ISSLER$^1$,
                           James T.\ JENKINS$^2$,
                           Jim N.\ McELWAINE$^3$}

\affiliation{%
$^1$Norwegian Geotechnical Institute, Oslo, Norway\\
E-mail: di@ngi.no\\
$^2$School of Civil and Environmental Engineering, Cornell University,
Ithaca, NY, U.S.A.\\
$^3$Department of Earth Sciences, Durham University, Durham, UK%
}

\abstract{In a series of papers, Bartelt and co-workers developed novel
  snow-avalanche models in which \emph{random kinetic energy} $R_K$ (a.k.a.\
  granular temperature) is a key concept. The earliest models were for a
  single, constant density layer, using a Voellmy model but with
  $R_K$-dependent friction parameters. This was then extended to variable
  density, and finally a suspension layer (powder-snow cloud) was added. 
  The physical basis and mathematical formulation of these models is
  critically reviewed here, with the following main findings:
  (i) Key assumptions in the original RKE model differ substantially from
  established results on dense granular flows; in particular, the effective
  friction coefficient decreases to zero with velocity in the RKE model.
  (ii) In the variable-density model, non-canonical interpretation of the
  energy balance leads to a third-order evolution equation for the flow depth
  or density, whereas the stated assumptions imply a first-order equation.
  (iii) The model for the suspension layer neglects gravity and disregards
  well established theoretical and experimental results on particulate
  gravity currents.
  Some options for improving these aspects are discussed.}

\maketitle

\section{Introduction}
\label{sec:intro}

More than half a century after \citet{Vo55} published his heuristic
bed-friction law for snow avalanches,
\begin{equation}
  \label{eq:Voellmy}
  \bss = -\frac{\ubb}{|\ubb|}
  \left( \mu \sigma_n + \frac{g}{\xi} \rho |\ubb|^2 \right),
\end{equation}
it is still at the heart of most avalanche flow models that are used in
practical applications like hazard mapping, the design of protection dams
or dimensioning of buildings \citep[][\eg]{VoKl99,SaZw04,ChKoBa10}. $\bss$
is the bed shear stress, $\sigma_n$ the basal bed-normal stress, $\ubb$
the depth-averaged flow velocity, $\rho$ the flow density, $\mu$ the
dry-friction coefficient, and $g/\xi$ a dimensionless \emph{turbulent}
drag coefficient, with $g$ the gravitational acceleration. Calibration
work, \eg\ \citep{BuFr80,Gr98,BlEgNa02}, and accumulated experience from
practical application of the model led to recommended parameter pairs
($\mu,g/\xi$) that vary strongly with avalanche size, assumed avalanche
frequency, terrain form and altitude \citep{SaBuGu90,ChKoBa10}. Typical
ranges are $0.15 < \mu < 0.5$ and $0.003 < g/\xi < 0.03$, with very
particular events needing even lower values of $\mu$ and/or higher values
of $\xi$ (see also \citep{An12}). Recent high-resolution observations also
show that such models cannot describe avalanche motion with constant
coefficients \citep{KoMESoAsBr16}. This indicates that the model does not
correctly capture important physical mechanisms at work in snow
avalanches, chief among them flow-regime transitions that alter the
mechanisms generating friction, and erosion and entrainment of the snow
cover. (\citet{IsGa08} illustrate the effect of flow-regime transitions.
For the importance of erosion, see \eg\ \citep{EgDe05,MaRoHuMaFaLu10}.)
A consequence is that such models cannot make predictions from a
priori measurable data such as snow characteristics and topography (often
called class-1 predictions in the engineering literature). Instead,
extensive calibration using past events is needed for each region in
which one wishes to apply them.

A number of attempts have been made to replace Eq.~(\ref{eq:Voellmy}) by a
formulation closer to the physics of granular media, \textit{e.g.}
\citep{SaGu85,Gu89,NoIrSc87,IsGa08}, but they have only met with partial
success and are rarely used in practice. Perhaps the most comprehensive and
ambitious of these attempts is by Bartelt and co-workers, who set out to
modify the Voellmy-type model RAMMS \citep{ChKoBa10} with features suggested
by the theory of granular flows, centred on what these authors term
\emph{random kinetic energy}. These papers can be characterised briefly as
follows (we will refer to them henceforth as [I]--[VII]):
\begin{description}
\item[{[I]:}] \citet{BuBa09} [Production and decay of random kinetic energy
  in granular snow avalanches] introduce the notion of random kinetic
  energy (RKE), propose a balance equation for it and fit the model to
  experimentally observed velocity profiles. They also indicate an
  exponential dependence of Voellmy's friction parameter $\mu$ on RKE.
\item[{[II]:}] The RKE-dependence is extended to $g/\xi$ by \citet{BaBu10}
  [Frictional relaxation in avalanches], who also discuss a number of
  conceptual issues.
\item[{[III]:}] \citet{BaMeBu11} [Snow avalanche flow-regime transitions
  induced by mass and random kinetic energy fluxes] reduce the model
  sketched in [II] to a block model that can be described by ordinary
  differential equations and study its properties as a dynamical system
  (fixed points, stability, flow in phase space) in detail.
\item[{[IV:]}] In this paper [Modelling mass-dependent flow regime
  transitions to predict the stopping and depositional behaviour of snow
  avalanches], \citet{BaBuBuChMe12} formulate the model from [II] as a
  depth-averaged model, solve it numerically and test it against a number
  of full-scale experiments.
\item[{[V]:}] In order to account for avalanche volume expansion due to
  particle collisions, \citet{BuBa11} [Dispersive pressure and density
  variations in snow avalanches] develop equations for the vertical motion
  of the centre-of-mass of a control column in the avalanche under the
  action of what they call dispersive pressure.
\item[{[VI]:}] Extending the approach in [V], \citet{BuBa15} [An
  energy-based method to calculate streamwise density variations in snow
  avalanches] supplement the model from [IV] with three further conservation
  equations connected to dispersive pressure.
\item[{[VII]:}] \sloppy\citet{BaBuVVBu16} [Configurational energy and the
  formation of mixed flowing/powder snow and ice avalanches] equip the model
  introduced in [VI] with a second layer for the powder-snow cloud and
  propose that it is formed by intermittent ejection of a mixture of fine
  snow grains with air from the dense core.
\end{description}
Recently, \citet{BaBu16} applied their approach to debris flows. We chose
not to include that paper in the present discussion because \citet{IvGe16}
already provided a concise critique of the way the authors use the notions
of excess pore pressure and particle--fluid interactions.

These seven papers use non-standard terminology in many instances. To help
readers who read the original papers [I]--[VII] alongside the present
analysis, we will use that terminology most of the time. However, it may be
useful to establish correspondence with standard terminology in advance for
a few key terms:
\begin{description}
\item[Random kinetic energy (RKE):] `Fluctuation energy' or `granular
  temperature' are established notions to describe the same phenomenon.
  There is a potential pitfall in the definition given, \eg, in [VII],
  which reads ``\ldots the kinetic energy associated with all particle
  movements different from the mean velocity of the flow''. Apparently,
  the mean velocity is to be understood as the velocity averaged over both
  time and the flow depth. In shear flows, a large part of this energy is
  not random, but due to the non-uniformity of the mean velocity profile.
  In [I] and [II], no depth-averaging is involved and velocity profiles are
  explicitly discussed. There is also ambiguity since fluctuation energy
  can be present at the grain scale (granular temperature) or at an eddy
  scale (turbulent kinetic energy).
\item[Dispersive pressure:] This is taken to mean the excess of the
  slope-normal stress at the base over the slope-normal component of the
  depth-integrated weight; it is positive when the avalanche dilutes and
  negative when it contracts. Readers familiar with the literature on snow
  avalanche dynamics should note that the same term was used a quarter
  century earlier by \citet{NoIrSc87,NoIrSc89} to designate the (always
  positive) pressure due to particle collisions. \citet{IvGe16} also
  discuss the notion of dispersive pressure in the context of a follow-up
  paper by \citet{BaBu16} on debris-flow modelling.
\item[Configuration energy:] This is a central notion in [VI]. It is
  otherwise probably most often used in the context of atomic and many-body
  physics, where it describes the potential energy of the system due to the
  mutual interactions between its components (usually by means of
  electromagnetic forces). In the context of [VI] and [VII], it is simply
  the depth-integrated gravitational potential energy density of the
  avalanche core relative to densest random packing.
\item[Plumes:] In [VII], the authors use this term for puffs of the
  air-snow mixture that are ejected near the avalanche front. `Plume' has
  a standard, technical meaning in fluid dynamics, namely a column of one
  fluid moving through another. In the extensive literature on gravity
  currents, `plume' usually refers to the entire gravity current on an
  incline, \eg\ see \citep{Si87}. What the authors are describing
  is simply turbulent entrainment by eddies.
\end{description}

So far, these models have not been discussed in the literature by other
workers. Given the wide-spread use of RAMMS and the prospect of the proposed
extensions being moved into the production version, a critical assessment of
the physical foundation of the models and their mathematical implementation
is called for. Our analysis of the papers [I]--[VII] revealed that the model
assumptions have a number of important implications that appear not to have
been recognised earlier. Also, we found that a number of fundamental
equations in the earlier papers were tacitly corrected in later papers,
which can make the reading confusing at times. Yet, we believe there remain
problems with the proposed mechanisms, regarding both their physical
plausibility and their mathematical formulation, that need to be addressed
before the model can be applied to practical problems with confidence.

We will first discuss the concept and mathematical formulation of the base
model without density variation ([I]--[IV]) in Section~\ref{sec:RKE},
then scrutinise its extension to variable density ([V]--[VII]) in
Section~\ref{sec:variable_density}. We consider the problems posed by the
suspension layer (powder-snow cloud) in [VII] separately in
Section~\ref{sec:PSA} and conclude with suggestions for further work in
Section~\ref{sec:conclusion}. Technical details are relegated to the
appendices. Due to space constraints, the present paper cannot be completely
self-contained.  We reproduce key equations and attempt to summarize the
argumentation of the authors, but refer the readers to the original papers
for details and precise wording.

\section{The basic RKE model}
\label{sec:RKE}

There are significant differences in the mathematical formulation of
some key concepts between the papers [I]--[IV]. Our analysis in this
section will concentrate on a comparison between the Voellmy friction
law, the RKE-modified friction law (most succinctly proposed in [III]
and [IV]), and an example from the kinetic theory for granular
materials, adapted from \citep{JeAs99}. The papers [I] and [II]
contain several conceptual errors that were tacitly corrected in [IV];
as they do not affect the formulation of the model directly, but have
led to confusion among readers, we will briefly discuss them in
Appendix~\ref{sec:app_A}.

\textbf{Summary of model assumptions.}
Paper [IV] starts from the long-established depth-averaged balance
equations of mass and momentum. For the earliest formulation in
snow-avalanche dynamics, see \citep{Eg67,PlRaTa84}; \citet{PaFuPa86}
give a detailed derivation in a two-layer situation similar to the one
discussed in Sec.~\ref{sec:PSA}. In a coordinate system following
the terrain, these equations read as follows if terms due to
non-orthogonality and curvature \citep{GrWiHu99,Gr01,BoWe04} are
neglected:
\begin{eqnarray}
   \partial_t h + \nablab \cdot (h \ubb) \!&\!=\!&\! Q ,
										        \label{eq:mass_bal}	\\
   \partial_t (h\ubb) + \nablab \cdot (h \ubb\ubb) \!&\!=\!&\!
      - \nablab \left( \frac{1}{2} g_z h^2 \right) + h \bm{g}
                                               \nonumber			\\
      & &\! - \frac{\ubb}{|\ubb|}
              \left[\mu h g_z + \frac{g}{\xi} \ubb^2\right] ,
					       						\label{eq:mom_bal}
\end{eqnarray}
where $\ubb = (u,v)^{\mathrm{T}}$, $\nablab = (\partial_x, \partial_y)$ and
$\bm{g} = (g_x, g_y)^{\mathrm{T}}$ denote the depth-averaged mean flow
velocity, the two-dimensional gradient operator, and the slope-parallel
components of the gravitational acceleration vector, respectively. Finally,
$h$ is the flow depth, and $Q$ is the volumetric entrainment rate, which we
will not discuss here. Incidentally, Eq.~(\ref{eq:mom_bal}) assumes
$\overline{\bm{uu}} \approx \ubb^2$, \ie, a uniform velocity profile. This
is a poor approximation if RKE is important, as the velocity profiles
presented in [I] show.

The basic RKE model departs from standard Voellmy-type models like RAMMS in
that the friction parameters $\mu$ and $k$ in Eq.~ (\ref{eq:mom_bal}) are
postulated to depend on $R_K$, the RKE density due to fluctuations of the
particle velocities about their mean values, as
\begin{equation} \label{eq:mu_k_of_RK}
   \mu(R_K) = \mu_0 \ee^{-R_K/R_0} , \qquad
   \frac{g}{\xi(R_K)} = \frac{g}{\xi_0} \ee^{-R_K/R_0} .
\end{equation}
With $R_0$, a new parameter enters the model that needs to be determined
empirically. A further balance equation describing advection, production
and dissipation of RKE complements the balance equations shown above:
\begin{equation} \label{eq:RKE}
   \partial_t(R_K h) + \nablab \cdot (R_K h \ubb)
   = \alpha \bss \cdot \ubb - \beta R_K h ,
\end{equation}
where $\bss$ is the bed shear stress given by the second line in
(\ref{eq:mom_bal}), and $0 < \alpha < 1$, $\beta > 0$ are two
parameters. Crucial ingredients of the model are the production and
dissipation terms for $R_K$ as well as the postulated dependence of
the Voellmy friction parameters, Eq.~(\ref{eq:mu_k_of_RK}).

While the papers [I]--[IV] refer to the kinetic theory of granular flows,
they depart from one of its well-established results without mentioning
or justifying their choice: Both for dilute and dense granular flows (and
also for turbulent fluid flows), the dissipation rate is found to grow as
$R_K^{3/2}$ \citep{JeRi85,JeBe10}. This is known as Haff's law and is a
basic result of kinetic theory \citep{Ha83}. If dissipation is assumed to
grow only linearly with $R_K$ as in in Eq.~(\ref{eq:RKE}), much higher
equilibrium values of $R_K$ result for a given production rate. Equation
(\ref{eq:RKE}) contains another strong assumption for which no
justification is given, namely that a fixed fraction of the shear
dissipation rate is converted into RKE.

\textbf{Velocity dependence of the effective friction coefficient in
        the RKE model.}
From Equations~(\ref{eq:mu_k_of_RK}) and (\ref{eq:RKE}), one can deduce
the speed of steady, uniform flow for a given slope angle $\theta$ and
flow depth $h$, and thus the effective friction law: The left-hand side
of Eq.~(\ref{eq:RKE}) vanishes in steady, uniform flow so that
\begin{equation} \label{eq:RKE_equi}
  R_K^\infty
  = \frac{\alpha}{\beta} \frac{\bss  \cdot \ubb}{h}
  = \frac{\alpha\rho}{\beta h}
    \left[ \mu_0 g_z h + \frac{g}{\xi_0} \ubb^2 \right] |\ubb|
    \ee^{-R_K^\infty / R_0} .           
\end{equation}
For simplicity, let us define $U \equiv |\ubb|$, the non-dimensional
RKE $r \equiv R_K^\infty / R_0$, the velocity scale
\begin{equation} \label{eq:vel_scale}
   U_0 \equiv \frac{\beta R_0}{\alpha \mu_0 \rho g_z} ,
\end{equation}
and the effective Voellmy friction coefficient,
\begin{equation} \label{eq:mu_Voellmy_eff}
   \mu_\mathrm{eff}^\mathrm{V} \equiv c_\mathrm{V} \mu_0
         \equiv \left(1 + \frac{g U^2}{\xi_0 \mu_0 g_z h} \right) \mu_0 .
\end{equation}
This brings Eq.~(\ref{eq:RKE_equi}) into the form
\[ r(U,h) = c_\mathrm{V} \frac{U}{U_0} \,\ee^{-r} , \]
which is solved by
\[
   r(U,h) = W_0(c_\mathrm{V} U / U_0) .
\]
$W_0$ is the upper branch, defined on $(\ee^{-1}, \infty)$, of Lambert's $W$ function, which is the solution to $x = W(x) \exp(W(x))$. As $x \rightarrow \infty$, also $W(x) \rightarrow \infty$, but for $x > \ee$, $W(x) < \ln x$. Applying this to Eq.~(\ref{eq:mu_k_of_RK}) and carrying out a few algebraic manipulations involving the defining equation of $W(x)$, we can rewrite the effective friction coefficient of the RKE model in steady, uniform flow as
\begin{equation} \label{eq:mu_eff_RKE}
   \mu_\mathrm{eff}^\mathrm{RKE}(U,h)
   = \mu_0 \frac{U_0}{U}\, W_0(c_V U / U_0) .
\end{equation}

\textbf{Comparison with kinetic theory.}  It is interesting to compare
this heuristic bed-friction law to one derived using methods of the
kinetic theory for collisional grain flows. We consider an essentially
passive slab that is suspended and transported on a relatively thin
region of intensely sheared grains at its base \citep{JeAs99}. This
may represent a slab avalanche in a very early phase, when the slab is
only about to disintegrate and is gliding on the thin weak layer whose
collapse caused the avalanche to release. We do not propose this model
as a replacement for any other model, but have chosen it because it
describes the same flow configuration as the Voellmy model \citep{Sa93},
namely a deformable but essentially passive heap riding on a thin,
intensely sheared basal layer. Yet, the Jenkins--Askari model shows
distinctly different behaviour from both the original Voellmy model or
the basic RKE model of [IV].

Expressions for the shear stress, normal stress, energy flux, and rate
of collisional dissipation at the base result from detailed consideration
of the transfer of momentum and energy in particle collisions with a
bumpy, rigid boundary \citep{Ri88}. Their values are obtained as solutions
of boundary value problems for the random kinetic energy and average
particle velocity in the thin region of intense shearing at the base.
The resulting expression for the dynamic, or rate-dependent part, of the
ratio of the shear stress, $S$, and normal stress, $P$, at the base, when
added to the rate-independent part, $\mu_0$, gives the effective friction
coefficient \citep{JeAs99}:
\begin{equation} \label{eq:mu_JA}
   \mu_{\mathrm{eff}}^\mathrm{JA}
   = \mu_0 - \alpha \left(\frac{4g_z h}{U^2}\right)^{1/2}
     + \left[\alpha^2\frac{4g_z h}{U^2} + \frac{12+\pi}{5\pi}(1-e)\right]^{1/2}.
\end{equation}
Here, $\alpha$ measures the difference between slip working and
collisional dissipation at the boundaries; it is a function of the
boundary roughness and the coefficient of restitution, $e_w$, in a
collision between a flow sphere and the boundary. $e$ is the
coefficient of restitution in a collision between two particles in of
the flow. Note that the model is restricted to low and moderate velocities
so that the slab on top of the thin shear layer does not become strongly
agitated.

\begin{figure}
   \includegraphics[width=\columnwidth]{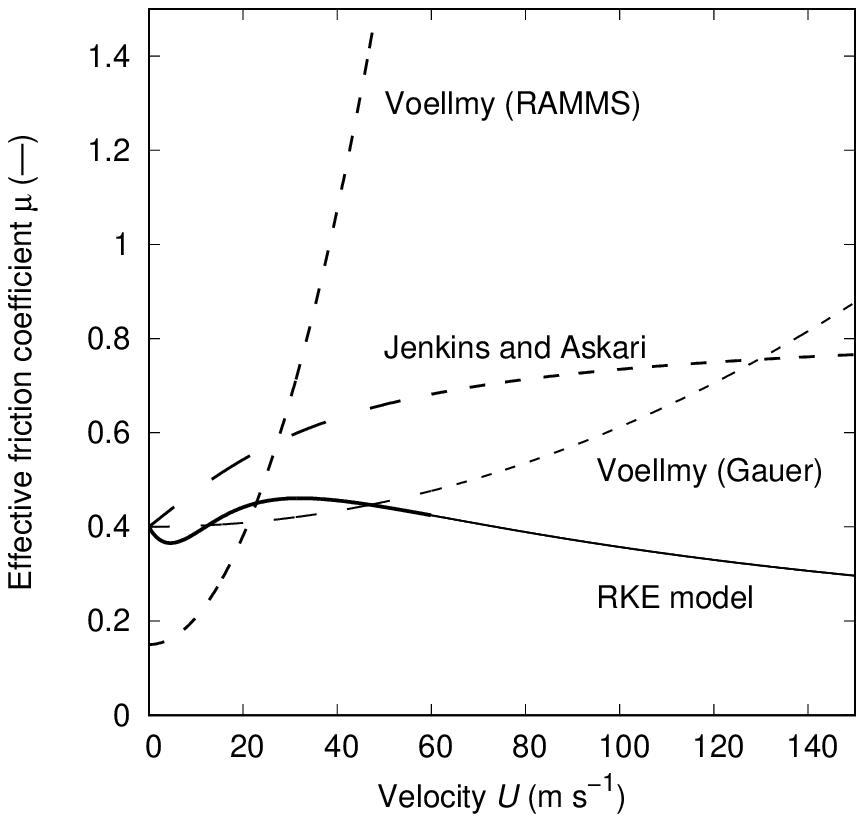}
   \caption{Comparison of the velocity dependence of the effective
     friction coefficient for an avalanche with flow depth 1\,m on a
     30\textdegree\ slope. (i) Voellmy model with traditional
     calibration \citep{ChKoBa10}, $\mu = 0.15$, $g/\xi = 0.0049$. (ii)
     Voellmy model with the calibration suggested by \citet{Ga14}, for
     an average slope angle $\beta \approx 30$\textdegree:
     $\mu_0 = 0.4$, $g/\xi = 0.00018$. (iii) \citet{JeAs99} model with
     $\mu_0 = 0.4$, $\nu_0 = 0.45$, $e = 0.85$, $e_w = 0.8$, and
     $\alpha = 0.463$. (iv) RKE model [IV] with $\mu_0 = 0.4$,
     $g/\xi_0 = 0.017$, $\alpha = 0.1$, $\beta = 0.5\,\mathrm{s}^{-1}$,
     $R_0 / \rho = 15\,\mathrm{m}^2\,\mathrm{s}^{-2}$.}
   \label{fig:mu_eff}
\end{figure}

Figure \ref{fig:mu_eff} compares the velocity dependence of the effective
friction coefficient of the original Voellmy friction law (in two
different calibrations), the kinetic theory, and the RKE-enhanced Voellmy
model. Note that these curves are valid only for the specific choice of
parameters. For all friction laws, we chose a slope angle of 30$^\circ$
and a flow depth of 1\,m. For the Voellmy model, we set $\mu = 0.15$ and
$g/\xi = 0.0049$ in the usual calibration, but $\mu = 0.4$ and
$g/\xi = 0.00018$ in the calibration proposed by \citet{Ga14}. For the
Jenkins--Askari model, we selected $\mu_0 = 0.4$, volume fraction at the
base $\nu_0 = 0.45$, and restitution coefficients $e = 0.85$ inside the
shear layer and $e_w = 0.80$ at its boundaries. Assuming a rough boundary
of smaller particles, we set $\alpha = 0.463$. For the RKE model,
$\mu_0 = 0.4$, $g/\xi_0 = 0.017$, $\alpha = 0.1$, $\beta =
0.5\,\mathrm{s}^{-1}$, and $R_0/\rho = 15\,\mathrm{m}^2\,\mathrm{s}^{-2}$
were assumed. With the traditional calibration of the Voellmy model, the
resistance is dominated by the velocity-dependent term already at speeds of
15--20\,m\,s$^{-1}$ so that the model generally underpredicts avalanche
velocities substantially. This defect is mitigated in Gauer's
calibration---at the expense of an almost linear dependence of $\mu$ on
the mean slope angle of the path and a linear dependence of $\xi$ on the
drop height \citep{Ga14}. The Jenkins--Askari model exhibits rapid growth
of $\mu_\mathrm{eff}$ at low velocity, but moderate growth at high
velocity, thus allowing the avalanche to reach high velocity even on
moderately steep slopes. Finally, the RKE model shows completely
different, non-monotonic behaviour, with $\mu_\mathrm{eff}$
\emph{decreasing} with increasing velocity above some threshold speed.
We caution, however, that Fig.~\ref{fig:mu_eff} takes the models beyond
their range of applicability. Within the realistic range 0--60\,\ms, a
constant value $\mu \approx 0.42$ and $g/\xi = 0$---corresponding to the
Savage--Hutter model \citep{SaHu89,SaHu91})---would give fairly similar
behaviour.

The Jenkins-Askari model predicts that stationary flows are only
possible in a limited range of slope angles (22--38\textdegree\ for the
choice of parameters in Fig.~\ref{fig:mu_eff}). The original Voellmy law
(\ref{eq:Voellmy}) clearly has a lower bound $\theta \geq \arctan \mu$,
but no upper bound because $\mu_\mathrm{eff}$ grows as the square of the
velocity. A stationary flow can thus be attained in arbitrarily steep
terrain. From Figure \ref{fig:mu_eff}, one can infer the following
scenario in the RKE model: An avalanche starting in sufficiently steep
terrain will quickly attain high velocity and low effective friction. If
the slope angle decreases in such a way that always $\tan\theta(x)
\gtrsim \mu_\mathrm{eff}^\mathrm{RKE}(U(x))$, the avalanche will
continuously accelerate while the slope angle tends to 0. Of course,
slopes on Earth are too short for this to happen, but the theoretical
possibility illustrates that Eq.~(\ref{eq:mu_k_of_RK}) is an extremely
strong assumption.

It has been known for a long time that the shear stress in rapidly
sheared granular materials increases with the square of the shear rate if
the volume is held constant. If the experiment is carried out at constant
normal stress (as in free-surface chute flows), the shear stress increases
also, but much more slowly due to the material expanding. For example,
experiments on a wide range of slope angles \citep{HoME12} show that the
effective friction always increases with velocity for a fixed flow rate.
Kinetic theory successfully predicts this behaviour.

It is conceivable that one sometimes can obtain satisfactory simulations
of observed events if completely different mechanisms have a similar
friction-reducing effect, \eg, lubrication by a thin water layer,
progressive comminution of snow particles, or excess pore pressure.
However, if the objective is to construct a physically founded model of
snow avalanche motion, it appears more promising to adopt the key results
from the theory of granular flows and to add such non-granular effects in
a more specific manner.

\textbf{Comparison with experiments.}  Paper [IV] points out that chute
experiments with snow \citep{PlBaKe07} show significantly smaller ratios
of basal shear stress $S$ to basal normal stress $N$ in the head of the
flows than at the snout and in the tail. The authors attribute this
difference to different levels of RKE and argue that, therefore, a separate
balance equation for RKE is needed. However, this conclusion does not
follow: The entire flow mass started from rest simultaneously, so its flow
depth, velocity and RKE should evolve in the same way unless some other
process differentiates between head and tail. Longitudinal normal stresses
reinforce gravity at the snout and counteract it in the tail, leading to a
significant decrease of the velocity from snout to tail \citep{PlBaKe07}.
Thus, a shear-rate-dependent rheology may also be able to capture the
observed variation in the ratio of shear stress to normal stress.

There is, however, a finding from the chute experiments \citep{PlBaKe07}
that deserves closer consideration: Independent of whether the flow is
stationary or not, $S/N$ is the effective bed friction coefficient,
$\mu_\mathrm{eff}$. The snow chute experiments indicate that
$\mu_\mathrm{eff}$ is higher in the slower tail than in the faster head,
while the snout, which moves at the same speed as the head, also has a
high value of $\mu_\mathrm{eff}$. This appears to be in stark contrast to
the findings of the granular--flow experiments and the granular rheology,
in which $\mu_\mathrm{eff}$ increases monotonically with velocity. It
should be illuminating to reanalyse the snow-chute experiments in terms of
granular rheology in order to confirm or refute the discrepancy with the
granular experiments. One may then try to relate the variation of
$\mu_\mathrm{eff}$ to, \eg, the dominant particle size at different
locations in the flow.

\textbf{Is a dynamical equation for RKE necessary?}  Another interesting
question is whether the extra balance equation for RKE is really
necessary. Certainly, it is needed in a complete theory of granular flows
because RKE (or granular temperature) may be produced at one location,
transported by advection and diffusion, and finally dissipated somewhere
else. Bartelt and co-workers emphasise that RKE is produced at the bed
interface and diffuses into the avalanche body. Much of the RKE is
certainly produced near the bed, where the shear rate is highest, but in
highly agitated flows with a Bagnold-type velocity profile, significant
RKE production occurs also inside the flow so that bed-normal diffusion of
RKE need not play a dominant role in the balance equation. In fact kinetic
theory shows that boundary effects decay exponentially, with a decay length
that is only a few particles long for dense inelastic flows. This means the
equilibrium granular temperature is slaved to the local shear rate with an
extremely rapid relaxation time. Moreover, the flow model is depth-averaged
and bed-normal RKE diffusion is not directly visible in the model; this
makes it much less compelling to use an extra equation for RKE unless the
dissipation coefficient $\beta$ is very small, which hardly can be the
case in a dense granular flow with rather inelastic collisions. Typical
values of the inverse RKE decay constant $1/\beta \gtrsim 1\,\mathrm{s}$
used in [IV] are short compared to the macroscopic time scale of avalanche
flow, \ie, the RKE is almost always close to its instantaneous equilibrium
value. In addition it is inconsistent to include a process with such rapid
relaxation whilst excluding processes with much slower relaxation.

The RKE model, using the Voellmy friction law as a basis, inherits
from it the general disadvantage that it only specifies the basal
shear stress. If the shear stress inside the flowing material or the
normal stresses are needed, additional assumptions have to be
made. \citet{BaBuPl06} assume the Voellmy model for the basal shear
stress, a viscous-frictional model inside the flow and an exponential
decrease of the RKE with distance from the bed in order to approximate
velocity profiles measured at the test site Vall\'ee de la Sionne. In
particular, these rheological assumptions determine the bed-normal
stresses inside the flow. This approach was, however, abandoned in
subsequent papers.

\section{The variable-density model}
\label{sec:variable_density}

Paper [V] marks an important turning point relative to [I]--[IV] at the
conceptual level: It acknowledges---albeit somewhat ambiguously---that
RKE can do work expanding the flow in the bed-normal direction:
\begin{quote}
   \textit{%
   In general, the energy, $R$, is random in nature and therefore $R$
   cannot perform mechanical work. However, at the base of the avalanche
   a flux of $R$ is deflected by the running surface upwards into the
   segment [a bed-normal control volume across the depth of the avalanche]
   (\ldots). The granular burst is given by the flux, $\dot{R}$. This
   energy flux raises the center of mass, converting a random energy
   flux into potential energy (it performs mechanical work).}
\end{quote}
In fact, it is a fundamental property of agitated granular masses that
the random particle motion creates stresses inside the mass and at solid
boundaries. Where particle impacts move a boundary against an
externally applied force or where stress gradients inside the granular
flow accelerate a portion of the mass, the granular mass performs
mechanical work. This process is analogous to the conversion of internal
energy (heat), associated with random molecular motion, in a combustion
engine.

In [V], the emphasis is on the bed-normal motion of a column of avalanching
snow with constant mass hold-up; RKE is assumed given. Paper [VI] extends
this mass-point or infinite-slope model to a complete flow model, with a
dynamical equation for the RKE. In [VI], there are a few changes in the
equations for bed-normal motion (or, equivalently, density change) due to
the newly introduced concept of configuration energy, which is defined as
the difference in depth-integrated gravitational potential energy relative
to a completely settled configuration at the same location on the slope.

Buser and Bartelt decompose the bed-normal stress at the bed, $N^{(b)}$,
into the weight-induced part $N_g^{(b)} = \rho g_z h \equiv m g_z$ with
$m$ the constant mass hold-up (or mass per unit footprint area) in this
idealised situation, and the dispersive (or excess) pressure $N_K^{(b)}$,
which is proportional to the bed-normal acceleration of the centre-of-mass
position $k$. With slightly changed notation,
\[
   N_K^{(b)} = m \ddot{k} .	\tag{V.4}
\]
Assuming a uniform density profile, one may approximate $k = h/2$. At
this point, a constitutive equation specifying $N^{(b)}$ (or $N_K^{(b)}$)
in terms of the flow variables $h$ (or $k$), $\ubb$ and $R_K$ is needed.
In [V] and [VI], Buser and Bartelt pursue two different approaches, which we
will examine in turn.

\textbf{The approach of [V]: analogy with an ideal gas.}
In the text following (V.6), Buser and Bartelt postulate that the equation
of state of a granular snow avalanche is essentially equivalent to that
of an ideal gas, \ie, they assume
\begin{equation} \label{eq:EoS_ideal_gas}
   N^{(b)} h = \gamma R_K h
\end{equation}
in their notation. Next, they take a total time derivative of the
equation of motion (V.4) to arrive at (V.5) and then do the same with
the equation of state shown above. However, in doing so, they only keep
$\gamma \dot{R}_K h$ on the right-hand side and omit $\gamma R_K \dot{h}$:
\[
   \frac{\dd (N^{(b)} h)}{\dd t} = \gamma \dot{R}_K h . \tag{V.6}
\]
With $\dot{N}_g^{(b)} \equiv 0$ and $\dot{N}_K^{(b)} \equiv \dot{N}^{(b)}$,
this mathematical error leads to the third-order equation adopted in [V],
\[
   \dddot{k} + (g_z + \ddot{k}) \frac{\dot{k}}{k}
   = \frac{\gamma \dot{R}_K}{m} . \tag{V.7}
\]

Indeed, if one used Eq.~(\ref{eq:EoS_ideal_gas}) in (V.4) and set
$N_K^{(b)} = N^{(b)} + m g_z$ (note that $g_z < 0$), one would immediately
arrive at the second-order equation
\begin{equation} \label{eq:vertical_motion_KT}
   \ddot{k} = \gamma \frac{R_K}{m} + g_z .
\end{equation} 
As the avalanche expands perpendicularly to the bed, the RKE density $R_K$
tends to decrease both because the energy is distributed over a larger
volume and energy is expended in working against gravity.

A more detailed analysis of the analogy between thermodynamics and the
kinetic theory of granular materials suggests to replace the constant
$\gamma$ by a function of the particle volume fraction, $f(\nu)$. This
function depends on the details of the particular approximation one
chooses. Using $\nu = \nu_0 k_0 / k$, we can express $f(\nu)$ in terms
of $k$ as $\tilde{f}(k) = f(\nu_0 k_0 / k)$ and replace $\gamma$ with
$\tilde{f}(k)$ in Eq.~(\ref{eq:vertical_motion_KT}).

\textbf{The approach of [VI]: configuration energy.}
Paper [VI] arrives at Eq.~(VI.25), which is essentially equivalent to
(V.7), without invoking analogy with an ideal gas or making an explicit
assumption for the constitutive equation; hence we need to discuss this
approach separately. As mentioned above, an important difference between
[V] and [VI] is the use of energy balances throughout [VI]. In particular,
the notion of configuration energy density (CED), $R_V$, is introduced.
In [VI], Sec.~3, it is defined as the gravitational potential energy per
unit volume, averaged over the depth of a bed-normal column, relative to
the configuration with maximum random packing at the same location.
Buser and Bartelt introduce the production rate of CED, $\dot{P}_V$, and
postulate it to be a fixed fraction $\gamma$ of the net production rate
of the sum of RKE and CED.

The critical step of their derivation is described between Eqs.~(VI.20)
and (VI.21), which we quote here (note that the authors switched notation
from $h$ to $V_\Phi$, but we will use $h$ in what follows):
\begin{quote}
	\textit{The total work done per unit time by the normal pressure at the
	bottom of the avalanche $N$ [\ldots] must be in balance with the
	total working of the particle interactions per unit volume. We have
	termed this change in potential energy as the configurational energy
	production $\dot{P}_V$. Therefore, the total change in the volume is
	\[
       \partial_t (N V_\Phi) = \dot{P}_V V_\Phi . \tag{VI.21}
    \]}
\end{quote}
Equation (VI.21) above is identical with (V.6) if one identifies the
coefficients $\gamma$ in [V] and [VI] with each other, extends $R_K$ in
(V.6) to $R = R_K + R_V = R_K / (1-\gamma)$, and replaces the partial time
derivative $\partial_t$ with the advective derivative $\Dd_t$ to account
for the translational motion of the control volume. We find this text
passage somewhat ambiguous, but understand it as making the following
interrelated statements:
(i) At the mesoscopic level, particle interactions in the avalanche do
mechanical work, which manifests itself as (and is quantitatively equal
to) the mechanical work done by the basal pressure at the macroscopic
level.
(ii) The rate of mechanical work done by the pressure due to particle
collisions must be equal to the rate of change of the configuration
energy.
(iii) The work rate of the bed pressure is $\Dd_t(Nh)$.
The paragraphs below will analyze these three statements in detail. Note
that we will henceforth write $P_V$ instead of $\dot{P}_V$ used by Bartelt
and co-workers. Their notation suggests that the production rate is the
time derivative of some other quantity, which, however, is never introduced
and would have no other meaning than the total of the RKE produced in an
advected unit volume---in particular, it is not the RKE.

\textbf{Pressure, work rate and configuration energy.}
The first issue to note is that, contrary to the statement (i) above, the
pressure at the bottom does not do mechanical work because the bed-normal
velocity $w(0)$ vanishes at that boundary. At the top surface, $w(h) \neq
0$, but the pressure (relative to atmospheric pressure) vanishes. The
mechanical work is being done inside the mass, where $w(z) \neq 0$ and
$N(z) \neq 0$. This is not merely a semantic point because it leads to
extra coefficients in the expression for the work rate that are missing on
the left-hand side of Eq.~(VI.21). We will detail this after mentioning
the remaining issues.

Second, statement (ii) neglects the change of kinetic energy associated
with bed-normal motion that unsteady expansion necessarily induces. A
third issue is intertwined with this one: The text as well as Eq.~(VI.21)
set the change in configuration potential energy equal to the work done
by the total pressure $N = N_g + N_K$ (statements (ii) and (iii)
combined). However, only $N_g(z)$ contributes to the change of CED,
whereas the gradient of dispersive pressure, $\partial_z N_K$, accelerates
the avalanching mass in the $z$-direction and changes the corresponding
contribution to the kinetic energy density $\mathcal{K}_z \equiv
\overline{\rho w^2}/(2h)$.

Fourth, Eq.~(VI.21) stipulates that the work rate of the pressure is
$\Dd_t (Nh)$; no further explanation for this assertion is given. The
expression $\Dd_t (Nh)$ contains the three terms $\dot{N}_K h$,
$N_g \dot{h}$ and $N_K \dot{h}$ ($\dot{N}_g h$ is zero if there is no net
mass flux into or out of the control volume). If Eq.~(VI.21) were true,
pressure would do mechanical work whenever it increases, even if $h$ is
held constant. A simple example is heating of a gas in a rigid container:
The gas does not do mechanical work in this process, but its capacity to
do so increases.

\begin{figure}
  \includegraphics[width=0.4\textwidth]{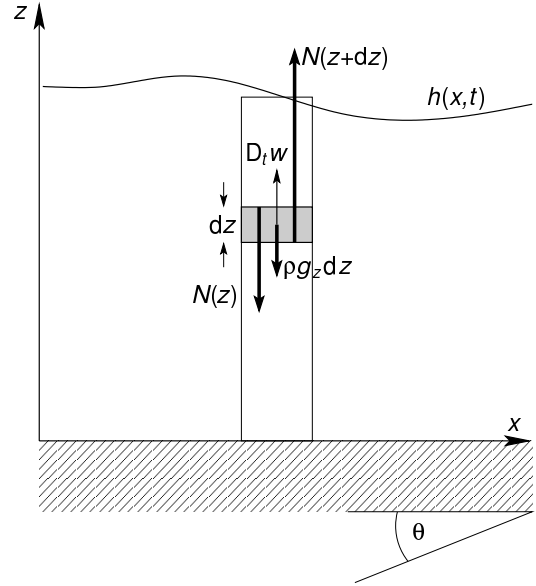}
  \caption{Schematic representation of an infinitesimally thin column of
    an avalanche on a plane inclined at an angle $\theta$ and an
    infinitesimal control volume within that column. Only the forces
    relevant for the bed-normal motion are indicated.}
  \label{fig:control_volume}
\end{figure}

In Appendix~\ref{sec:app_B}, we will compare the model [VI] to the
general balance equations for mass, momentum and fluctuation energy in a
depth-averaged flow model. It may be instructive, however, first to
illustrate the issues mentioned above in a simple, quasi-one-dimensional
setting, disregarding variations in the $x$ and $y$-directions. We
assume, however, that shearing motion in these directions produces RKE.
The control volume under consideration and the stresses and body forces
in the $z$-direction are schematically represented in
Fig.~\ref{fig:control_volume}. We first consider the momentum balance and
then turn attention to the energy balance.

The momentum balance for a thin slope-parallel slice of thickness $\dd z$
at height $z$ is given by
\begin{equation} \label{eq:momzbal}
   \partial_t (\rho w) + \partial_z (\rho w^2) = \rho g_z - \partial_z N.
\end{equation}
Integrating this equation over $z$ from 0 to $\infty$, using the boundary
conditions $\rho(\infty) = w(0) = N(\infty) = 0$ and introducing the mass
hold-up $m \equiv \int \rho(z) \dd z$ as well as the centre-of-mass
velocity $\bar{w} \equiv \int \rho w \dd z$, one obtains
\begin{equation} \label{eq:momzbal_int}
   m \dot{\bar{w}} = m g_z + N^{(b)} .
\end{equation}
where $N^{(b)}$ is the value of $N$ at the bed. If one assumes the density
to be uniform across the flow depth but variable in time, $w(z,t) =
2 \bar{w}(t) z/h$ for kinematic reasons. Similarly, $\dot{w}(z,t) =
2 \dot{\bar{w}}(t) z/h$ in the Lagrangean sense. The dynamics, governed
by Eq.~(\ref{eq:momzbal}), then requires the pressure to vary with
$\zeta \equiv z/h$ as
\begin{equation} \label{eq:press_profile}
   N(z,t) = m \left[ -(1-\zeta) g_z + (1 - \zeta^2) \dot{\bar{w}}(t)
              \right] .
\end{equation}
The kinetic and configuration energies are given by
\[
   \bar{\mathcal{K}}h + R_V h
   \equiv h \int_0^1 \frac{\rho}{2} w^2 \dd\zeta
          - \int_0^1 \rho g_z h\zeta \dd\zeta + m k_0 g_z .
\]
If the density profile is uniform, these energies become
\begin{equation} \label{eq:pot_kin_energies_uniform}
   \bar{\mathcal{K}}h = \frac{4}{3} \cdot \frac{1}{2} m \bar{w}^2
   \quad \mathrm{and} \quad
   R_V h = -m \left(\frac{h}{2} - k_0\right) g_z
\end{equation}

We can calculate the depth-integrated work rate $\dot{W}$ from the
well-known expression for the work rate per unit volume,
$\dot{\mathcal{W}}$, of the internal (Cauchy) stresses in a continuous
medium:
\begin{equation} \label{eq:work_rate_Cauchy}
   \dot{\mathcal{W}} = \sigma_{ij} D_{ij}
      \equiv \sigma_{ij} \frac{1}{2}(\partial_j u_i + \partial_i u_j).
\end{equation}
In our case and with the assumption that the density profile is uniform,
the strain rate has only one non-zero component, given by $D_{zz} =
\partial_z w = 2\bar{w}/h$; the stress tensor has the corresponding
component $\sigma_{zz}(z,t) = -N(z,t)$. A straightforward calculation of
$\dot{W}$ using Eq.~(\ref{eq:press_profile}) then yields
\begin{equation} \label{eq:work_rate_integ}
   \dot{W} = \frac{4}{3} m\bar{w}\dot{\bar{w}} - m \bar{w} g_z .
\end{equation}
Comparison with Eq.~(\ref{eq:pot_kin_energies_uniform}) immediately shows
$\dot{W} = \Dd_t[(\bar{\mathcal{K}} + R_V) h]$, as it should be.

Now, if $\Dd_t(N^{(b)} h)$ indeed is the work rate of pressure, directly
evaluating it must give the same result as in
Eq.~(\ref{eq:work_rate_integ}). From Eq.~(\ref{eq:press_profile}) and the
assumed uniform density profile, we obtain
\begin{equation} \label{eq:VI26_eval}
   \Dd_t[N^{(b)} h]
   = 2 m \left[ \Dd_t(k\ddot{k}) - \bar{w}g_z \right]. 
\end{equation}
The right-hand sides of Eqs.\ (\ref{eq:work_rate_integ}) and
(\ref{eq:VI26_eval}) differ significantly. Equation
(\ref{eq:work_rate_integ}) is derived directly from the principles of
continuum mechanics, with the only additional assumption of $w(z,t)$
being linear in $z$. Equation (\ref{eq:VI26_eval}) follows from the
postulate (VI.26) combined with the definition (VI.3), the balance
equation (VI.8), the postulate (VI.12) and assumed linearity of $w(z,t)$.
This leaves two alternatives: Either $D_t(N^{(b)}h)$ is not a correct
expression for the depth-integrated work rate of the pressure, or
Eq.~(VI.21) must be considered an implicit constitutive assumption for
the granular pressure. The first alternative has far-reaching
consequences: (VI.21) must be abandoned and the basis for the
mathematical development in the rest of Secs.~5 and~6 of [VI] is
invalidated. Instead, one has to adopt Eq.~(\ref{eq:work_rate_integ}) and
state a suitable constitutive equation for the granular pressure as a
function of the flow variables, \eg, $N = N(R_K)$.

Now consider the second alternative: In this case, $\Dd_t(N^{(b)}h)$ is
\emph{not} the total work rate of the granular pressure. In order for
$\Dd_t(N^{(b)}h)$ to be equal to $\Dd_t(R_V h)$, the equation
$\Dd_t(N^{(b)}h) = \dot{W} - \Dd_t(\bar{\mathcal{K}}h)$ must hold. In the
case of linear $w(z,t)$ with $h \equiv 2k$ and $\bar{w} \equiv \dot{k}$,
this leads to
\[
   \Dd_t(N^{(b)}h) \equiv 2\Dd_t(N^{(b)}k) = -m \dot{k} g_z .
\]
If $m$ and $g_z$ are constant along the avalanche path, we can perform
the time integration (more precisely, the integration along the
characteristic line of the control volume) easily and obtain
\[
   N^{(b)}(t) = -\frac{1}{2} m g_z + \mathrm{cst.}
\]
If the avalanche is at rest (and its depth is $2k_0$), $N^{(b)}$ must
equal the weight, thus $\mathrm{cst.} = -\frac{1}{2} m g_z$. However,
this leaves no room for a dynamical evolution of the flow depth, and we
conclude that the second alternative is not viable.

In our opinion, the most immediate solution of this dilemma in the energy
formalism is to explicitly state a constitutive equation for $N^{(b)}$,
to express the work rate of the granular pressure in terms
of $N^{(b)}$ as
\begin{align}
   \dot{W} &= 2\bar{w}
              \int_0^1 \left[ N_g^{(b)} (1-\zeta)
                              + (N^{(b)} + m g_z) (1-\zeta^2)
                       \right] \dd\zeta					\nonumber \\
           &= \left[-m g_z + \frac{4}{3} (N^{(b)} + m g_z)\right] \bar{w}
\end{align}
and to set it equal to the rate of change of kinetic and configuration
energy. Using Eq.~(\ref{eq:pot_kin_energies_uniform}), dividing by
$\bar{w}$ and rearranging terms, we arrive at
\begin{equation}
   m\dot{\bar{w}} = N^{(b)} + m g_z .
\end{equation}
Not surprisingly, this is the same as Eq.~(\ref{eq:vertical_motion_KT})
if one makes the constitutive assumption $N^{(b)} = \gamma R_K$. We note
that such modification of the extended RKE model also requires modifying
the RKE balance equation to properly account for the conversion of RKE
to kinetic and configuration energy:
\begin{align}
   \partial_t(R_K h) &+ \bm{\nabla}\cdot(R_K h \ubb) \nonumber \\
       &= \alpha \dot{W}_f - \beta_K R_K h
          - \left(\frac{4}{3}\gamma R_K + \frac{1}{3}m g_z\right) w .
                                               \label{eq:R_K_balance}
\end{align}
This will be discussed further when comparing the extended RKE model with
the general form of the balance equations for mass, momentum and RKE in
Appendix~\ref{sec:app_B}.

\textbf{First-order equation for the flow depth implied by the model
assumptions.}
Our fifth remark is that the modelling assumptions put forth in the
\emph{text} of [VI] together with the energy partitioning postulate
(VI.12) imply a simple, first-order evolution equation for $h$. As the
cited text from [VI] states, Eq.~(VI.21) is to describe the (advected)
rate of change of CED, $\Dd_t R_V \equiv \partial_t R_V
+ \ubb\cdot\bm{\nabla} R_V$. Its right-hand side is the production rate
of CED, $P_V$, integrated over the flow depth. Buser and Bartelt
postulate the following equations:
\begin{align*}
   P_V &= \gamma P ,                              \tag{VI.12} \\
   P h &= \alpha \dot{W}_f^{xy} - \beta_K R_K h , \tag{VI.8}  \\
   \dot{W}_f^{xy} &= \bss  \cdot \ubb_\parallel . \tag{VI.6}
\end{align*}
In an avalanche starting at rest, $R_K(0) = R_V(0) = 0$. Due to(VI.12),
$P_K = (1-\gamma) P$ so that the advected rates of change of $R_K$ and
$R_V$ are proportional. This immediately leads to
\begin{equation} \label{eq:R_V_prop_R_K}
   (R_V h)|_{\bm{x},t} = \frac{\gamma}{1-\gamma} (R_K h)|_{\bm{x},t}
                       = \gamma (Rh)|_{\bm{x},t} . 
\end{equation}
The system of the first four equations defined by (VI.30),
(VI.37)--(VI.39), is closed by the assumptions (VI.34)--(VI.36) for
$\bm{S}_b$ together with Eq.~(\ref{eq:R_V_prop_R_K}) and specified values
for $\alpha$, $\beta_K$ and $\gamma$. Moreover, throughout [VI] (and
[VII]), the authors assume in addition that the density profile is
uniform. Then the depth-integrated change rate of the CED can be
expressed straightforwardly in terms of $\dot{h}$ and the mass hold-up:
The integral of the potential energy over $z$ relative to the reference
configuration is
\begin{equation} \label{eq:int_pot_energy}
   R_V h = \bar{\rho} g_z h (k-k_0)
         = m g_z \left( \frac{h}{2}-k_0 \right) .
\end{equation}
We will disregard terrain curvature and entrainment for the sake of
simplicity. In the Lagrangian point-of-view, $\Dd_t m = 0$. From this,
we obtain directly
\begin{equation} \label{eq:first_order_h_evol}
   \Dd_t h = \frac{2\gamma}{m g_z}
             \left[ \alpha \bss \cdot\ubb - \beta_K R_K h \right] .
\end{equation}
This first-order differential equation describes relaxation of $h$ to a
(quasi-)steady-state value governed by the RKE. This enslavement of $R_V$
and $h$ is a direct consequence of the very strong assumption (VI.12).
We emphasise that this first-order evolution equation for $h$ is implied
by the model assumptions stated in [VI] and that the additional three
equations in (VI.30), (VI.37)--(VI.39) at best are superfluous. If one
desires a model with more complicated dynamics in the $z$-direction, one
has to replace the assumption (VI.12) by a weaker one that does not
enslave $R_V h$.

Closer examination reveals that Eq.~(\ref{eq:first_order_h_evol}) does
not strictly conserve energy: According to the assumptions made in [VI],
part of the frictional work is directly converted to heat, the rest to
$Rh = (R_K + R_V) h$, where $R_V$ is potential energy and $R_K$ is
fluctuation energy, \ie, $\langle w'^2 \rangle = 0$ for a suitably defined
time or ensemble average $\langle.\rangle$. However, if the density
changes, $\Dd_t h \neq 0$ and $\overline{\rho w^2} > 0$, thus there is a
contribution from the slope-normal expansion to the total kinetic energy
that is not accounted for in the modified model. As mentioned earlier, it
might be more natural to include this non-random part of the kinetic
energy in the CED; if one does so, energy is conserved. However,
Eq.~(\ref{eq:int_pot_energy}) is then no longer valid and a separate
evolution equation for the flow depth must be constructed. As explained
above, Eqs.~(VI.5)--(VI.7) are not suitable for this.

As in the case of the basic RKE model, an analysis of the time scales
associated with different processes---in this case, relaxation of the
depth-averaged velocity, the RKE, the velocity profile and the density
to their steady-state values---is required for a consistent approximation
of avalanche flow. We cannot pursue this question further here, but the
tight coupling between $h$, $\ubb$ and $R_K$ revealed by
Eq.~(\ref{eq:first_order_h_evol}), the rapid relaxation of $R_K$ and our
experience from studying alternative snow avalanche models all suggest
that algebraic equations for $R_K$ and $h$ instead of differential ones
would produce a simpler, yet more consistent and equally accurate model.

\textbf{Mathematical formulation of the flow model.}
A final remark concerns [VI], Sec.~6, where all model equations are
reformulated for implementation in a numerical code. The procedure for
doing so is well-known and correctly applied for the conserved
quantities mass, momentum and RKE in Eqs.~(VI.30)--(VI.34). Sec.~VI.5
then states that the three first-order evolution equations
(VI.27)--(VI.29) for $k$, $w$ and $N_K$ (obtained from the erroneous
third-order equation (VI.17)) are extended to include advection, which
indeed is necessary. However, Eqs.~(VI.37)--(VI.39) present these
equations in conservation form, which does not follow from the advected
form for non-conserved quantities. For example, the difference between
the conservative and advective extensions of (VI.27) is
\[
   \partial_t k + \bm{\nabla}\cdot(k\ubb) - \mathrm{D}_t k
   = \bm{\nabla}\cdot\ubb .
\]
Clearly, the (two-dimensional) divergence $\bm{\nabla}\cdot\ubb$ does not
vanish identically.

\section{The mixed-avalanche model}
\label{sec:PSA}

\sloppy \textbf{General considerations.}  The notion of powder-snow
avalanche is somewhat fuzzy in the literature. To avoid ambiguity, we
will use the term ``mixed snow avalanche'' (MSA) for flows that
simultaneously feature three different flow regimes, namely dense flow
(DF), light (or intermediate-density or fluidised) flow (LF), and
suspension flow (SF) (see \cite{SoMELo15} for an in depth discussion of
these regions). The original RKE-extension of the Voellmy model
(Sec.~\ref{sec:RKE}) with constant density is, in principle,
applicable only to the DF, but is in practice used to model both the
DF and LF regimes. The variable-density model
(Sec.~\ref{sec:variable_density}) attempts to explicitly model
transitions between the DF and LF regimes in a single-layer model.  In
[VII], the SF regime is added to the model through a second layer.

Based on today's knowledge from observations and measurements, the LF regime
may be attained in small avalanches, but is typically more strongly developed
in larger avalanches. The SF regime will not be attained unless a
considerable part of the avalanche has reached the LF regime. Due to their
elevated velocity, parts of the avalanche in the LF regime will reach farther
than the parts in the DF regime, and the part in the SF regime may travel yet
farther (by several kilometres in extreme cases).

There is probably a smooth transition between flow regimes. This would
favour a mathematical description in terms of a multi-phase model (air
and snow particles of different sizes), where the density and the stress
tensor depend on the volumetric concentrations of the different
particle size classes and where air turbulence plays an important role
at low particle concentration. However, such a model would have to be
formulated as a genuine three-dimensional model and would at present
be poorly suited for practical applications. The different deposit
characteristics \citep{IsGaScKe96} of the DF and LF regimes as well as
some measurements with FMCW radar suggest that the three flow regimes
nevertheless may often be fairly distinct, with large density gradients
at their boundaries. This opens the way for models with several layers
corresponding to different flow regimes and with depth averaging applied
to each layer.%
\footnote{We note that recent detailed measurements in large MSA's at the
Vall\'ee de la Sionne test site in Switzerland \citep{SoMELo15,KoMESoAsBr16}
suggest a more complex picture in which sudden, intermittent bursts of
rather large and high-density volumes of snow particles play an important
role---perhaps not unlike horse-shoe vortices detaching from the bottom
surface in turbulent flows. If confirmed, these measurements may question
the traditional approach of modelling avalanches in terms of continuum
models with slowly varying, depth-averaged fields.}
Virtually all models proposed so far follow this path.

Considered in isolation, the SF is a turbulent suspension of small
snow grains in air and a sub-type of particulate gravity currents
\citep{Si87}.  The volume concentration of the grains is usually very
low ($\ll 0.1$) so that grain interactions are not important, though
due to the high density of the grains they carry most of the
momentum.
The excess density of the mixture over the air is referred to as the
\emph{buoyancy} of the current. The density is not constant as on the
upper surface ambient air is entrained by turbulence and on the lower
surface snow can be lost due to particle settling or gained due to
entrainment. All models must therefore have at least three equations
for momentum, air mass and snow mass or --- equivalently and more
commonly --- buoyancy and volume. In its initial and final stages, the
SF is in the Boussinesq regime where the average density is almost the
same as the ambient air density, but it is far in the non-Boussinesq
regime when fully developed.

There is a large body of experimental, theoretical and numerical work
on density and particulate gravity currents in a variety of idealised
settings. It is an important question which of these results remain valid
in the case of MSAs and must be taken into account in the modelling
of the SF regime. Density and turbidity currents in the laboratory are
typically produced from a dilute initial suspension and run over a
relatively smooth, non-erodible bed. In contrast, the SF in an MSA forms
at the front and on top of the highly agitated LF, and mass is exchanged
between the LF and the SF at a high rate. The particle concentration in
the SF layer typically being less than 0.01 and the particle settling
velocity less than 1\,m\,s$^{-1}$, the mechanism of particle suspension
inside the layer has to be the same as in other dilute particulate gravity
currents and the processes at the upper surface that govern entrainment
of ambient air have to follow the general laws observed in
high-Reynolds number jet flows, plumes or thermals.

A number of theoretical analyses determine the front velocity of gravity
currents by treating them as inviscid, energy-conserving flows without
explicitly taking into account turbulence \citep{Be68,ME05,NoDaStVeOl08}.
The applicability of the results from this approach appears a priori
questionable, but recent numerical studies nevertheless appear to confirm
it at least for flows without a dense undercurrent \citep{KoLSMEMe16}.
In all these situations, one finds that the bed shear stress is negligible
compared to the effect of ambient-fluid entrainment along the upper
surface. It remains to be seen, however, whether this also holds true in
MSAs, where there presumably is a pronounced vertical density gradient
across the depth of the SF layer and where the surface of the DF/LF layer
beneath can be strongly agitated. Both these flow properties increase the
shear stress at the lower interface of the SF layer.

\textbf{Basic modelling assumptions in [VII].}  The approach proposed in
[VII] is a two-layer formulation: a lower layer of intermediate to high
(variable) density, consisting of large snow balls, fine snow grains
and air for the DF and LF regimes, and an upper layer of low (variable)
density containing air and fine snow grains in the SF regime. The air in
the SF layer is treated as incompressible and the layer depth and mass
per unit footprint area are used instead of air and snow mass. The
relative motion between snow grains and air is neglected so that a
single momentum balance equation is sufficient for the SF layer.
Perhaps surprisingly, the balance of turbulent energy is not considered
here, even though the concept of RKE is borrowed from the theory of
turbulence, and turbulence is instrumental in maintaining the snow grains
in suspension \citep{FuPa90,PaFuPa86}.

We need not discuss the left-hand sides of the balance equations further
because they have standard form. Paper [VII] postulates the following
source terms for the conservation equations for mass, $x$ and $y$-momentum
and volume of the suspension layer:
\[
\bm{G}_\Pi = \left(\!\!\!\begin{array}{c}
    \dot{M}\FP + \dot{M}\LP \\
    \dot{M}\FP u_\Phi - S_{\Pi x} \\
    \dot{M}\FP v_\Phi - S_{\Pi y} \\
    \dot{V}\FP + \dot{V}\LP
  \end{array}\!\!\!
\right) . \tag{VII.35}
\]
$\dot{V}\FP$ and $\dot{V}\LP$ are the volume fluxes from the dense
core ($\Phi$) and the ambient air ($\Lambda$) to the suspension layer
($\Pi$), and $\dot{M}\FP$ and $\dot{M}\LP$ are the associated mass
fluxes. $\bm{S}_\Pi$ denotes the surficial shear stress on the
suspension layer. In what follows, we will change the notation from
$\dot{M}\FP$, $\dot{M}\LP$, $\dot{V}\FP$, $\dot{V}\LP$ to $Q\FP$,
$Q\LP$, $W\FP$, $W\LP$ because these quantities are \emph{not}
advective derivatives of $M_\Phi$, $M_\Pi$, $h_\Phi$ and $h_\Pi$, as
the dot notation would imply.

\textbf{The role of gravity.}  A particular feature of the MSA model of
[VII] --- immediately apparent from the second and third component of
$\bm{G}_\Pi$ in (VII.35) --- is that gravity is neglected in the dynamics
of the SF layer, both as the driving force and as the cause of
sedimentation. This is different to every other model of gravity
currents and contradicts a large body of well-documented research
showing that gravity and air entrainment at the top surface are the
dominant terms in the momentum balance of density currents
\citep{Ho83,MeMEKn12}. Such an approximation might be justified when
describing the motion of jets of almost particle-free air ejected from
the avalanche, but this would be devoid of practical relevance.
Neglecting sedimentation is less grave unless one is interested in the
late run-out phase of the PSA, where the flow rarely does damage.

The missing processes are easily introduced into the source term
(VII.35):
\begin{equation} \label{eq:PSA_source_term}
   \bm{G}_\Pi'
   = \left(\!\!\!\begin{array}{c}
        Q\FP - Q\PF + Q\LP \\
        (M_\Pi-h_\Pi\rho_a)\bm{g} + (Q\FP-Q\PF)\bm{u}_i - \bm{S}_\Pi \\
        W\FP - W\PF + W\LP
                 \end{array}\!\!\!
   \right) ,
\end{equation}
where $\rho_a$ is the density of the ambient air. The buoyancy term (the
left-most term in the middle row of Eq.~(\ref{eq:PSA_source_term})) is
present in all earlier PSA models we are aware of. The grain-borne shear
stress (middle term in the second row) represents the momentum flux from
one layer to the other due to the mass flux. It is the product of the
net entrainment rate (entrainment minus sedimentation) and the
(slope-parallel) velocity at the interface, $\bm{u}_i$. The latter needs
to be modelled as a function of $\ubb_\Phi$, $\ubb_\Pi$ and the densities
in the two layers, but we will not go further into this question. The
relation between mass and volume loss of the PSA cloud due to settling of
snow grains is $Q\PF = \rho_i W\PF$, assuming that all particles in the
PSA cloud are snow grains with the density of ice, $\rho_i$. A candidate
model for the sedimentation rate is the one used by \citet{PaFuPa86},
\begin{equation} \label{eq:PSA_sedimentation}
   Q\PF \approx c_b \left(\frac{M_\Pi}{h_\Pi} - \rho_a\right)
                w_s \cos\theta ,
\end{equation}
where $M_\Pi / h_\Pi$ is the depth-averaged PSA density, $w_s$ the
average settling velocity of the snow particles, and $\theta$ the
local slope angle. $c_b$, the ratio of bottom snow concentration to
depth-averaged concentration, needs to be assumed, the most plausible
values being in the range 3--10.

\textbf{Air entrainment and drag on the suspension layer.}
Paper [VII] does not specify the air entrainment rate $Q\LP$ appearing
in (VII.35) and in the first row of Eq.~(\ref{eq:PSA_source_term}).
This entrainment rate has been measured repeatedly in inclined plumes and
particulate gravity currents since the pioneering experiments by
\citet{ElTu59}. It is well understood by now that it is governed by the
Richardson number \citep{Tu73}. (\citet{TuMEAn07} provide a summary of
these ideas applied to avalanches.) The Richardson number is the ratio
of the potential energy to the kinetic energy of a parcel of fluid. For
an entire layer, one defines the \emph{bulk} Richardson number,
\begin{equation} \label{eq:Richardson}
   \mathrm{Ri} = \frac{(\rho-\rho_a) g h \cos\theta}{\rho_a u^2},
\end{equation}
where $\theta$ is the angle between the slope normal and vertical.
This implies that the entrainment rate depends both on the slope
inclination, as shown experimentally \textit{e.g.} by \citet{ElTu59}
and \citet{BeOl91}, and on the density difference.

Based on laboratory experiments, \citet{Tu86} proposed the following
formula for the entrainment coefficient:
\begin{equation} \label{eq:E_of_Ri_Turner}
   E(\mathrm{Ri})
   = \left\{ \begin{array}{l@{\quad}l}
                \displaystyle
                \frac{0.08 - 0.1\, \mathrm{Ri}}{1 + 5\, \mathrm{Ri}} ,
                    & \mathrm{Ri} < 0.8 , \\[1em]
                0,  & \mathrm{Ri} \geq 0.8 .
             \end{array}
     \right.
\end{equation} 
\citet{An04a} more recently fitted unpublished data of Beghin by the
function
\begin{equation} \label{eq:KSBalphav}
   E(\mathrm{Ri})
   = \left\{ \begin{array}{ll}
                \mathrm{e}^{-\lambda \mathrm{Ri}^2},
                    & \mathrm{Ri} \leq 1,        \vspace{0.5em} \\
                \mathrm{e}^{-\lambda}/\mathrm{Ri},
                    & \mathrm{Ri} > 1,  
             \end{array}
     \right.
\end{equation}
where $\lambda = 1.6$. This provides an even better closure for the
volume and air-mass balances
\begin{equation} \label{eq:air_entrainment}
   W\LP = E(\mathrm{Ri}) |\ubb_\Pi|
   \quad\textrm{and}\quad
   Q\LP = \rho_a W\LP.
\end{equation}

Bartelt and co-workers assume the sum of the shear stresses on the upper
and lower interfaces of the SF layer to be proportional to the square of
the cloud velocity, $\bm{u}_\Pi^2$, and the cloud density, $\rho_\Pi$,
writing
\[
   \bm{S}_\Pi = -\frac{\ubb_\Pi}{|\ubb_\Pi|} \frac{g}{\xi_\Pi}
                 \rho_\Pi \ubb_\Pi^2 .     \tag{VII.37}
\]
The drag coefficient $\xi_\Pi$ is considered a constant to be selected
by the user. The authors state that the dominant contribution to the drag
is from air entrainment at the upper boundary.

There are two issues with these assumptions: (i) If indeed air
entrainment is the dominant contribution to the retarding forces on the
SF layer, $\bm{S}_\Pi \approx 0$ would result because entrainment of
ambient air at rest does not remove \emph{momentum} from the SF layer,
but distributes it over an increasing mass and thus decelerates the
flow. To see this, multiply the mass balance equation
\[ \partial_t(h\rhob) + \nablab\cdot(h\rhob\ubb) = Q \]
by $\ubb$ and subtract it from the momentum balance equation
\[
   \partial_t(h\rhob\ubb) + \nablab\cdot(h\rhob\ubb\ubb)
   = (\rhob-\rho_a) h \bm{g} - \nablab (h \bar{p}) .
\]
This produces the equation of motion
\[
   h\rhob \Dd_t \ubb = (\rhob-\rho_a) h \bm{g} - \nablab(h\bar{p}) - Q\ubb,
\]
which correctly features the decelerating force due to the
acceleration of the ingested mass if this mass originally is at rest.
(ii) Drag at the lower boundary should depend, not on the SF layer
velocity, but on the difference between SF layer and DF/LF layer
velocities. It is thus seen that the model presented in [VII] effectively
assumes the interfacial shear stress between the DF/LF layer and the SF
layer to be given by Eq.~(VII.37) whereas the entrainment function $W\LP$
is not specified in the paper. Replacing $\ubb_\Pi^2$ by
$(\ubb_\Pi - \ubb_\Phi)^2$ in (VII.37) and specifying $W\LP$ and $Q\LP$
according to Eq.~(\ref{eq:air_entrainment}) with either
(\ref{eq:E_of_Ri_Turner}) of (\ref{eq:KSBalphav}) would then provide a
physically consistent, but rather crude closure, provided the opposite
term $+\bm{S}_\Pi$ is added to the momentum balance equation of the
DF/LF layer: The modified Eq.~(VII.37) does not take into account that
the interfacial shear stress will depend strongly on the densities in
both layers. 

The authors correctly note that air entrainment at the upper interface
is the main source of resistance for the powder-snow cloud. However,
there are also pressure forces acting, though the distinction between
pressure drag and turbulent entrainment is not as straightforward as
it first appears. For pressure to have a net retarding force there
must be a separation region behind an object and a turbulent wake,
otherwise we have D'Alembert's paradox that there is no drag. Thus the
momentum transfer from the pressure drag goes into a momentum deficit
in the turbulent wake. This turbulent mixing region contains snow and
if we regard it as part of the SF layer, then correctly accounting
for the momentum balance here means that we do not need to directly
consider the pressure drag. A more detailed discussion and quantitative
analysis from direct numerical simulation is contained in
\citep{KoLSMEMe16}.

This is correct if we consider the momentum balance over a large
region containing the front, but if we wish for a more detailed model
this can be improved. The pressure distribution along the SF layer
surface is obtained by solving an elliptic potential flow problem in
the ambient fluid. At the nose of the avalanche there will be a
positive (with respect to the background pressure) stagnation
pressure, but moving back along the top surface the pressure decreases
linearly and may become negative \citep{ME05}. Perhaps the best method
would be to solve this potential flow problem using a boundary element
method \citep{NoDaStVeOl08}. This approach has not yet been directly
applied to geophysical flows (see \citep{DBElIsHaBrLi04} for a
simplistic approximation). In addition, the shallowness approximation
is violated since the front angle will be approximately 60 degrees
\citep{ME05}. Instead the more usual approach is to apply a front
condition by imposing a constant Froude number, thus setting the front
velocity as a function of the front height.

In addition, added-mass effects (which give additional drag when an
avalanche is accelerating) may also be relevant for powder-snow clouds
of low density \citep{TuMEAn07}.

\textbf{Formation of the suspension layer.}
The authors of [VII] interpret the cleft-and-lobe structure of MSA fronts
as firm evidence for what they term blow-out. This proposed mechanism for
SF layer formation can be summarised as a consequence of a hypothetical
``breather'' mode in the motion of the dense core: The latter would expand
periodically (normal to the slope) due to dispersive pressure and then
contract again. During expansion, air would be ingested and mixed with
fine snow dust; when the large particles fall down again, the nascent
powder-snow cloud would be expelled from the core as plumes (or rather
puffs in conventional terminology), forming the observed lobes and clefts.

Paper [VII] specifies the boundary fluxes for the suspension layer as (in
our notation)
\begin{align*}
   W\LF &= 2 w_\Phi \Theta(w_\Phi) ,            \tag{VII.24} \\
   Q\LF &= \rho_{\Pi_0} W\FP \Theta(w_\Phi) ,   \tag{VII.25} \\
   W\FP &= 2 w_\Phi \Theta(-w_\Phi) ,           \tag{VII.26} \\
   Q\FP &= \rho_{\Pi_0} W\FP \Theta(-w_\Phi) ,  \tag{VII.27}
\end{align*}
where the Heaviside distribution $\Theta(x)$ is 1 for $x < 0$ and 0
otherwise. Precisely speaking, (VII.24) and (VII.25) refer to the
avalanche core, and as such (VII.24) is merely a kinematic
statement. Together they describe the first half of one cycle in the
``breather'' mode that is invoked here. The second half of the cycle
transfers mass and momentum from the core to the SF layer and
is described by (VII.26) and (VII.27).

As stated above and by (VII.24), the core entrains air (without ice
dust) at the rate $W\LF$. In contrast, (VII.25) states that the core
entrains mass from the ambient air at the rate $Q\LF = \rho_{\Pi_0}
W\LF$, \ie, air laden with ice dust. This is likely a
typographical error, and $\rho_{\Pi_0}$ should be replaced by
$\rho_a$. Even so, $Q\LF$ does not appear in (VII.34) and (VII.35),
while the terms $W\LP$ and $Q\LP$ appearing in (VII.35) are not
explicitly defined in [VII]. Thus, the first, second, third and fifth
components in Eq.~(VII.34) should be amended to read
\begin{equation} \label{eq:G_Phi_corr}
   \bm{G}_\Phi 
   = \left(\!\!
        \begin{array}{c}
           Q_{\Sigma\rightarrow\Phi} - Q\FP + Q\LF \\
           G_x - S_{\Phi x} - (Q\FP - Q\PF) u_{i x} + S_{\Pi x}    \\
           G_y - S_{\Phi y} - (Q\FP - Q\PF) u_{i y} + S_{\Pi y}    \\
           \vdots \\
           -h_\Phi \bm{\nabla}\cdot\bm{u}_\Phi + 2 w_\Phi \\
           \vdots
        \end{array}\!\!
     \right) .
\end{equation}
The first term of the fifth component corrects for the conservation form
of the left-hand side of the balance equation (VII.28); see
Sec.~\ref{sec:variable_density} for further necessary modifications of
(VII.32) and (VII.34).

Besides these technical issues, the proposed SF formation mechanism
presents two conceptual problems. First, lobes and clefts cannot be
considered evidence for this mechanism. They form inevitably as
instabilities in all types of highly turbulent gravity flows, with or
without suspended particles, on completely flat and smooth surfaces
\citep{Si87}. Even if there is no dense underflow and the proposed
``blow-out'' mechanism cannot be operative, the same type of structure
is observed. Engulfing of ambient air by large eddies (similar in size
to the flow depth) is the main entrainment mechanism in turbulent
gravity currents and may indeed lead to the impression that dense jets
are ejected from the flow. In MSAs, the density gradients are expected
to be large except in the late stage; this will accentuate the
impression of jets shooting out. To make this proposal more than
speculation, one ought to show that the conventional mechanisms fail
to reproduce central aspects of MSA behaviour or, ideally, perform
measurements at the interface of the core and the suspension layer.

Another crucial aspect of the ``blow-out'' mechanism is not addressed in
[VI] or [VII], but in [V]: The core must undergo some kind of
intermittency or oscillatory behaviour with a rather large amplitude
to eject large volumes of snow--air mixture. Such behaviour indeed
arose in the model described in [V] since the second-order equation
for the vertical velocity derived in that paper has no damping
term. However, as we show in Sec.~\ref{sec:variable_density}, that
equation lacks a proper physical basis.

Finally, we note that the ``blow-out'' mechanism as described in [VII]
requires that there is no suspension layer above the core during the
first (expansion) half of the cycle; otherwise, the core could not
entrain ``pure'' ambient air, as is stated before (VII.24) and
(VII.25). This means that the puff of snow-air mixture that is
ejected during core contraction must immediately move away from the
core segment that ejected it before the next expansion begins. This
may work for the front, but core segments behind the nose will
typically have a puff above them when they expand, and thus
``swallow'' the suspension layer rather than ingesting ambient
air. There is no indication in [VII] of how the model prevents the
core from repeatedly incorporating what it has just expelled.

\section{Concluding remarks}
\label{sec:conclusion}

There can be little doubt that the general thrust of the work
described in [I]--[VII] is promising and will lead to more realistic
avalanche models over time. However, the preceding sections revealed
that the models in their present form have shortcomings both in their
physical foundations and their mathematical formulation that need to
be dealt with. Our main findings can be summarised as follows:
\begin{enumerate}
\item In sharp contrast with all other models in practical use today,
  the effective friction coefficient of the RKE-extended Voellmy model
  \emph{decreases} with speed rather than increasing (provided the RKE
  is reasonably close to its equilibrium value). This friction law
  should therefore be regarded as heuristic, and its predictions
  should be compared to detailed measurements and, \eg, the
  Jenkins-Askari model. It will not be suitable as the foundation of a
  comprehensive, physics-based theoretical model unless one can show
  it to be a controlled approximation to a more physical model.
\item Inclusion of density changes adds a considerable amount of
  complexity to an avalanche model. Papers [V]--[VII] attempt to
  circumvent some of it by adopting an energy-based approach rather
  than a momentum balance equation in the $z$-direction. For closure,
  [VI] and [VII] assume that the energy supply to the configuration
  energy density (CED) is a fixed fraction of the net production rate
  of the sum of CED and RKE. A critical issue is the expression for the
  work rate of pressure, with $\Dd_t (N h)$ used instead of $\frac{1}{2}
  N_g \Dd_t h$ in Equation (VI.21). This leads
  to a spurious third-order equation for the flow depth in [V]--[VII],
  in lieu of the first-order evolution equation that results if the
  stated framework is applied correctly. Additionally it is
  inconsistent to model such a rapid process by an evolution equation
  when an anelastic type algebraic closure is much more suitable.
\item If one abandons the simplistic assumption for the energy supply
  rate to the CED, a more realistic description of the
  variable-density system becomes possible. We conjecture that a
  physically consistent and realistic, yet relatively simple model
  results if one assumes a linear relation between RKE and total
  bottom pressure. Additionally, the RKE balance equation must be
  extended to account for the two-way coupling between RKE and
  CED. This system and possibly additional options need to be studied
  further in order to find a physically sound and practical model.
\item The balance equations of the SF layer in [VII] are in
  contradiction to all other models and firmly established experimental
  results on dilute gravity mass flows. The source terms should include
  gravity, particle settling/entrainment and entrainment of ambient air.
  The important questions which are as yet unclear or controversial are
  the following: Which parameterisation of the   entrainment rate as a
  function of Richardson number is most   appropriate? What is the form
  of the density profile? What is the form of the velocity profile? Is
  there a significant interfacial shear stress between the DF/LF layer
  and the SF layer other than the momentum flux associated with mass
  exchange?
\item The proposed mechanism for generating a PSA from a dense avalanche
  is novel, but highly speculative. It depends on a ``breather mode''
  being excited in the dense flow, for which evidence is at best
  inconclusive. There is need for further work comparing the mechanism
  proposed in [VII] to more conventional alternatives.
\end{enumerate}

We expect that the necessary changes can be incorporated in the model
without major difficulties and that they will simplify the equations
for the density-changing DF/LF layer considerably. There is a wide
spectrum of alternative formulations, in particular with regard to
the DF rheology and the generation mechanism for the SF layer, that
are worth exploring. In particular, the early work [I,II] on fitting
velocity profiles measured in full-scale experiments holds promise of
finding a consistent and experimentally verified rheology to replace
the heuristic RKE-modified Voellmy friction law.

\section*{Acknowledgments}

Reviewing this paper required an extraordinary amount of effort as it
effectively amounted to reviewing papers [I]--[VII] as well. We express
our sincere gratitude to C.~Ancey and an anonymous referee for their
insightful and constructive reviews and to N.~Eckert, S.~H.~Faria and
J.~C.~Cogley for their editorial advice. We are grateful for the
hospitality of the Max Planck Institute for the Physics of Complex
Systems in Dresden, Germany, where work on this paper started during the
geoflow16 workshop. DI acknowledges correspondence from Perry Bartelt
that clarified some questions related to paper [VI]. Part of DI's work
was funded by the grant for snow avalanche research from the Norwegian
Department of Oil and Energy, administrated by the Norwegian Water
Resources and Energy Directorate. JJ acknowledges travel support associated
with his visit to the Max Planck Program geoflow16 received from US NSF
award 1619768 to Cornell University.

\appendix

\section{Further remarks on the RKE model}
\label{sec:app_A}

The papers [I], [II] and [VI] invoke a number of concepts that look
plausible and innocuous at first, but warrant a more detailed discussion.
In this appendix, we collect issues that do not directly affect the model
equations.

\textbf{Viscous shear vs.\ inelastic collisions.}
Bartelt and Buser distinguish between viscous shear work and inelastic
collisions associated with RKE. The first notion stems from a macroscopic
description of the granular material, while the second notion corresponds
to a microscopic viewpoint. The kinetic theory of granular materials shows
in a precise mathematical way how particle collisions give rise to a close
analogue of viscosity in fluids \citep{JeSa83}. As in the theory of
turbulence, correlations of fluctuation velocities create a contribution to
the stress tensor in a granular assembly. This implies that one cannot
separate these notions from one another, as is done in [I,II].

\textbf{Work done by random particle motion.}
Equations (I.5) and (I.6) stipulate that the forces arising from the
random motion of particles average to zero because of their randomness
and thus RKE cannot be converted to kinetic or potential energy. In
papers [V]--[VII], this statement is tacitly revoked and $\gamma R_K$ is
effectively used as the dispersive pressure. But where precisely lies the
flaw in the argument of [I]? One can trace the problem to the stipulation
that random motion produces random forces that average to zero. However,
forces are exerted by one system on another. If system A is a granular
assembly and we disregard static electricity, A can exert a force on some
other system B only if A and B are in contact with each other. System B
would typically be a container wall, creating a boundary for the granular
assembly. This causes an asymmetry: Particles approaching the right-hand
side wall have a wall-normal velocity component $v_\perp > 0$, but after
the collision, $v'_\perp < 0$. The \emph{force} on the wall depends, not on
the average of $v_\perp$ and $v'_\perp$, but on $v_\perp - v'_\perp$, which
necessarily is larger than 0 due to the presence of the boundary. If the
particel collisions with the wall are strong and frequent enough, they will
push the container wall outward and do mechanical work.

\textbf{Energy balances.}
The energy balance (I.19) (or its equivalent forms (I.7) and (I.11)) looks
deceptively simple and straightforward:
\[
    \dot{R} + \dot{\Phi} + \dot{K} = \dot{W}_g - \dot{W}_f.  \tag{I.19}
\]
However, [I] emphasizes that $\dot{W}_f$ is always negative because the
friction force opposes the direction of motion. \emph{Subtracting} a
negative work rate from the avalanche energy would therefore increase that
energy, thus the sign of this term must be changed. We will henceforth
consider the equation with corrected sign. To emphasize the conservative
character of the gravitational force in contrast to the dissipative nature
of friction, we will apply (I.9), $\dot{U} = -\dot{W}_g$, in reverse and
use the gravitational potential energy instead of the gravitational work
rate. We thus discuss the equation
\[
    \dot{R} + \dot{\Phi} + \dot{K} + \dot{U} = -\dot{W}_f.  \tag{I.19'}
\]

The authors apply this energy balance in the framework of a depth-averaged
flow model. Moreover, the shear is assumed to be concentrated in a very
thin bottom layer, \ie, one assumes the velocity profile to be uniform
inside the avalanche and the slip velocity to be equal to the
(depth-averaged) flow velocity. Let us therefore test (I.19') by
considering a block of mass $m$ sliding on a horizontal surface, with
friction coefficient $\mu$ and initial velocity $u_0$. The equation of
motion is readily integrated:
\[ u(t) = u_0 - \mu g t \]
and gives the kinetic energy
\[
   K(t) = \frac{m}{2} u_0^2
          - \mu m g \left(u_0 t - \frac{1}{2} g t^2 \right) .
\]
The time-dependent term on the right-hand side exactly equals the work
done on the block by the external friction force,
\[ W_f(t) = \int_0^{s(t)} (-\mu m g)\, \dd s , \]
as $s(t) = u_0 t - \mu g t^2/2$. Thus we find $K(t) = K_0 + W_f(t)$.
Since there is no gravitational work in this case, this corresponds to
(I.19'), but with $\dot{R} + \dot{\Phi} = 0$. However, the frictional
work surely has been converted into heat ($\dot{\Phi} > 0$), so why does
it not show up in the balance equation? The answer is that the heat is
not generated inside the sliding block but at the boundary and (I.19')
lacks a term describing the heat flux across the boundary of the block.
The correct form of Eq.~(I.19) would therefore be
\begin{equation}
   \dot{K} + \dot{U} + \dot{R} + \dot{\Phi} = \dot{W}_f + Q_a,
\end{equation}
where $Q_a$ is the sum of the granular and thermal heat fluxes into the
avalanche, integrated over the control volume surface.

The equation of motion implies $\dot{K} + \dot{U} = \dot{W}_f$, thus we
get $\dot{R} + \dot{\Phi} = Q_a$, but we do not know the value of $Q_a$.
We can obtain some qualitative insight if we consider an (infinitesimally)
thin control volume along the interface, in which all the shear is
concentrated. Friction converts kinetic energy of the sliding block into
heat and RKE inside the shear layer at a rate $-\dot{W}_f > 0$. With its
infinitesimal volume, the shear-layer has, however, only a vanishingly
small capacity for storing this energy. This means that the total heat
and RKE flux out of the shear layer into the avalanche, $Q_a$, and the
snow cover, $Q_s$, must equal $-\dot{W}_f$. Clearly, $Q_s > 0$ in a snow
avalanche so that $0 < Q_a < -\dot{W}_f = \dot{R} + \dot{\Phi}$.
Comparing with (I.17),
\begin{equation}
  \dot{R}=-\alpha\dot{W}_f-\beta R,\tag{I.17}
\end{equation}
one sees that Bartelt and Buser assume $Q_s \approx 0$, $\dot{R} = \alpha Q
\ \beta R$, and $\dot{\Phi} = (1-\alpha) Q + \beta R$. With these
assumptions, the balance equation (I.19') reduces to
\begin{equation} \label{eq:total_energy_bal_corrected}
   \dot{R} + \dot{\Phi} + \dot{K} + \dot{U} \approx 0 .
\end{equation}
In contrast, the original Eq.~(I.19') (after correcting the sign error)
would imply that the snow cover absorbs all the frictional work. The
preceding analysis also applies to a flow with internal shear with a
few modifications.

In Sec.~7 of [VI], Buser and Bartelt attempt to show that their equation
system conserves energy. To this end, they split the (non-random) kinetic
energy into two components, defined as $K^{xy} \equiv
\overline{\rho\bm{u}^2}/2$ and $K^z\equiv \overline{\rho w^2}/2$, and state
the following balance equations:
\begin{align}
\dot{K}^{xy} &= \dot{W}_g^{xy} - \dot{W}_f^{xy} ,         \tag{VI.40} \\
\dot{K}^z    &= \Dd_t(R_V h) - \dot{W}_g^z - \dot{W}_f^z. \tag{VI.41}
\end{align}
Among multiple issues connected with (VI.41), we mention the following:
(i) The kinetic energy balances should be derived directly from the
momentum balance equations. In doing so, Eq.~(VI.40) would receive a
contribution from the (slope-parallel) gradient of normal stresses, and
Eq.~(VI.41) would be supplemented by a contribution due to dispersive
pressure.
(ii) The model is not fully closed in the sense that evaluating
$\dot{W}_f^z = \int_0^h \nablab\cdot\bm{S}(z)\, \dd z$ would require
constitutive expressions for the shear stresses $\bm{S}(z)$ across the
flow depth. The Voellmy-type bed-friction law provides only the bed shear
stress, $\bm{S}_b$.
(iii) Equation (VI.41) should contain either the rate of change of
potential energy, $\Dd_t(R_V h)$, or the work rate of gravity,
$-\dot{W}_g^z$, but not both. Gravity being a conservative force, the
change of potential energy is the opposite of the work done by gravity,
thus $\Dd_t(R_V h) = +\dot{W}_g^z$ with the sign convention of Eq.~(VI.5).
When this is taken into account, (VI.41) degenerates to $\dot{K}_z =
-\dot{W}_f^z$. As mentioned above, this relation lacks the main term,
namely the contribution from the dispersive pressure gradient.

\section{Comparison with the general balance equations for mass, momentum
    and fluctuation energy}
\label{sec:app_B}

Further insight into the significance of the constitutive assumptions in
the density-changing RKE model can be obtained by comparing it to the
general depth-averaged balance equations for mass, momentum and fluctuation
energy, of which it has to be a special instance if it is to be consistent.
For simplicity, consider flow down a straight, rigid incline at an angle
$\theta$ to the horizontal. We take $x$ in the flow direction, $y$
horizontal in the sliding plane and $z$ normal to the incline, with origin
at the base and positive upward, $x_\alpha = (x,y,z)^T$, $\rho$ the average
mass density of the grains, and $u_\alpha = (u,v,w)^T$ the components of
the ensemble-averaged grain velocity. We will first state the equations for
a general 3D flow and then discard the variations along the $x$ and
$y$-directions to make the comparison simpler.

From the general principles of fluid mechanics, the mass balance equation
must take the local form
\begin{equation}
\partial_t \rho + \partial_\alpha(\rho u_\alpha) = 0 .
\end{equation}
(We use tensor notation here to emphasise that these equations are
three-dimensional and switch to vector notation after depth-averaging.)
Take $\sigma_{\alpha\beta}$ to be the components of particle stress and
$f_\alpha = g (\sin\theta,0,-\cos\theta)^T$ the components of external
force per unit mass, with $g$ the gravitational acceleration. Then the
local balance of linear momentum is
\begin{equation}
\partial_t(\rho u_\alpha) + \partial_\beta(\rho u_\alpha u_\beta)
= \partial_\beta \sigma_{\alpha\beta} + \rho f_\alpha .
\end{equation}
With $\mathcal{K} \equiv (1/2)\rho u_\alpha u_\alpha$, the balance of mechanical
energy reads
\begin{equation}
\partial_t\mathcal{K} + \partial_\beta (\mathcal{K} u_\beta)
= \partial_\beta (u_\alpha \sigma_{\alpha\beta})
- \sigma_{\alpha\beta} \partial_\beta u_\alpha + \rho u_\alpha f_\alpha .
\end{equation}

The granular temperature, $T$, is defined as one-third of the mean square
of the particle velocity fluctuations and thus relates to $R_K$ as
$(3/2)T \equiv R_K$. Let $q_\alpha$ be the components of the flux of
fluctuation energy, and $\varGamma$ the rate of collisional dissipation
per unit volume. $R_K$ then has to obey the following
advection--diffusion--dissipation equation:
\begin{equation}
\partial_t(\rho R_K) + \partial_\alpha (\rho R_K u_\alpha)
= -\partial_\alpha q_\alpha + \sigma_{\alpha\beta} \partial_\beta u_\alpha
- \varGamma .
\end{equation}

Next, we average over $z$ between the bed at $z = 0$ and the surface
at $z = h(\bm{x},t)$. We use the notation $\bm{u} \equiv (u,v)^T$,
$u_\alpha \equiv (\bm{u}, w)^T$, $\partial_\alpha \equiv (\bm{\nabla},
\partial_z)^T$. The 3D stress tensor decomposes into the 2D tensor
$\sigma_{ab} \equiv \bm{\sigma}$, the 2D vector $\sigma_{az} = \sigma_{za}
\equiv \bm{S}$, and the 2D scalar $\sigma_{zz}$, with $a,b \in
\{x,y\}$. $\bss $ is the bed shear stress. For any field
$\psi(\bm{x},z,t)$, the depth average can be written as
$h \bar{\psi}(\bm{x},t) \equiv \int_0^{h(\bm{x},t)} \psi(\bm{x},z,t) \dd z$.
Leibniz's rule, \textit{e.g.}, $\partial_t \int_0^h \psi(\bm{x},z,t) \dd z
= \int_0^h \partial_t \psi(\bm{x},z,t) \dd z + \psi (\bm{x},h,t)
\partial_t h(\bm{x},t)$, and the kinematic boundary condition,
\begin{equation} \label{eq:kin_BC}
\partial_t h(\bm{x},t) + \bm{u}(\bm{x},h,t)\cdot\bm{\nabla} h(\bm{x},t)
= w(\bm{x},h,t) ,
\end{equation}
are repeatedly used together with the boundary conditions
$w(\bm{x},0,t) = 0$ and $\sigma_{\alpha\beta}(\bm{x},h,t) = 0$. For
simplicity, we assume the bed to be non-erodible and the density to be
constant with depth. In this case the height $h$ is a useful variable.
When the density varies strongly and the upper edge may not be well
defined, it is better to work with mass holdup $m = h\rhob$ and the
centre of mass, $Z = \overline{Z\rho}/\overline{\rho}$. We can rewrite
the kinematic boundary condition as a volume balance equation. Thus the
system is governed by five balance equations for, respectively, the
volume, the mass, the linear momenta in the flow plane and normal to the
bed, and the fluctuation energy:
\newlength{\colone} \newlength{\coltwo} \newlength{\colthr}
\settowidth{\colone}{$\partial_t(h\overline{\rho R_K})$}
\settowidth{\coltwo}{$\bm{\nabla}\cdot(h \overline{\rho R_K \bm{u}} - h\bar{\bm{q}})$}
\settowidth{\colthr}{$q_z|_0 - h \bar{\varGamma} + \cdots$}
\begin{align}
    \makebox[\colone][l]{$\partial_t h$}
        & + \makebox[\coltwo][l]{$\nablab \cdot(h\overline{\bm{u}})$}
            & = \makebox[\colthr][l]
                        {$h \bm{\nabla} \cdot \overline{\bm{u}} + \overline{w}$}
                                                    \label{eq:FD_VC} \\[0.5em]
    \makebox[\colone][l]{$\partial_t(h\bar{\rho})$}
        & + \makebox[\coltwo][l]{$\bm{\nabla}\cdot(h \overline{\rho \bm{u}})$}
            & = \makebox[\colthr][l]{$0$}
                                            \label{eq:FD_mass_bal} \\[0.5em]
    \makebox[\colone][l]{$\partial_t(h\overline{\rho \bm{u}})$}
        & + \makebox[\coltwo][l]
                    {$\bm{\nabla}\cdot(h\overline{\rho \bm{uu}}
                        - \bar{\bm{\sigma}})$}
            & = \makebox[\colthr][l]{$\bss  + h \bar{\rho}\bm{f}$}
                                            \label{eq:FD_xmom_bal} \\[0.5em]
    \makebox[\colone][l]{$\partial_t(h\overline{\rho w})$}
        & + \makebox[\coltwo][l]
                    {$\bm{\nabla}\cdot(h\overline{\rho w\bm{u}}
                        - h\bar{\bm{S}})$}
            & = \makebox[\colthr][l]{$-\sigma_{zz}|_0 + h\bar{\rho}f_z$}
                                            \label{eq:FD_zmom_bal} \\[0.5em]
    \partial_t(h\overline{\rho R_K})
        & + \bm{\nabla}\cdot(h \overline{\rho R_K \bm{u}} - h\bar{\bm{q}})
            & = \makebox[\colthr][l]{$q_z|_0 - h \bar{\varGamma} + \cdots$}
                                            \nonumber \\
    \makebox[\colone][l]
            {$\quad\cdots + h\left[\,\overline{\bm{\sigma}:\bm{\nabla}\bm{u}}
                                    + \overline{\sigma_{zz} \partial_z w}
                                    + \overline{\bm{S}\cdot\partial_z \bm{u}}
                                    + \overline{\bm{S}\cdot\bm{\nabla} w} \,
                              \right] .$}
        &
            &                               \label{eq:FD_FE_bal}
\end{align}
In addition, one must specify expressions relating the average of products
of fields to the product of averages, constitutive equations relating the
stresses $\bar{\bm{\sigma}}$, $\bm{S}$, $\sigma_{zz}$ to the fields $h$,
$\rhob$, $\ubb$, $\bar{w}$ and $\bar{R}_K$, and boundary conditions for the
fields and stresses.
Within the stated framework, these equations are general. Note that the
height equation (\ref{eq:FD_VC}), mass balance (\ref{eq:FD_mass_bal}) and
the $z$-momentum balance (\ref{eq:FD_zmom_bal}) need to be used jointly to
determine the flow depth and the density. Simplifying assumptions are
needed to close the equations and to make them tractable. However, any
approximations have to be compatible with the general structure of
Eqs.~(\ref{eq:FD_VC})--(\ref{eq:FD_FE_bal}). Equation~(\ref{eq:FD_VC}) can
be thought of in at least three different ways: Firstly, as we have
written it here, as a volume balance equation; secondly as a
kinematic equation $\partial_t h + \overline{\bm{u}}\cdot\bm{\nabla} h =
w_h$; and thirdly as an equation for the centre of mass $h/2$. This last
interpretation is the most general and most useful since it corresponds to
the gravitational potential energy and is well defined for any density
distribution including when there is no well defined upper surface.

Now we may compare these equations with the extended RKE model of [V]
and [VI]. The indices $\Phi$ and $\Sigma$ refer to the dense flow and
the snow cover, respectively. One readily identifies $M$ with
$h\bar{\rho}$. First, we focus on the equations for $M_\Phi$,
$M_\Phi u_\Phi$, $M_\Phi v_\Phi$ and $R h_\Phi$; we will discuss the
equations for $h_\Phi$, $M_\Phi w_\Phi$ and $N_K$ afterwards. The
left-hand sides of Eqs.~(\ref{eq:FD_VC})--(\ref{eq:FD_FE_bal})
agree with Eqs.~(VI.30) and (VI.37)--(VI.39) if one approximates the
depth-averages of products of fields by the products of the
depth-averaged fields, assuming uniform profiles. The source terms
proposed in [VI] are summarised by the first four rows of Eq.~(VI.39):
\[
\bm{G}_\Phi
= \left(\!\!\begin{array}{c}
\dot{M}_{\Sigma\rightarrow\Phi} \\
G_x - S_{\Phi x} \\
G_y - S_{\Phi y} \\
\alpha \bm{S}_\Phi \cdot \bm{u}_\Phi
- \beta_K (1-\gamma) R h_\Phi
\end{array}\!\!
\right)                                             \tag{VI.39}
\]
With the erosion rate set to 0, this becomes in our notation
\begin{equation} \label{eq: RKE_source_terms}
\bm{G}_\Phi = \left(\!\! \begin{array}{c}
0 \\[0.5ex]
h\rhob\bm{f} - \bss   \\[0.5ex]
\alpha \bss \cdot\ubb
- \beta_K (1-\gamma) h\overline{\rho R_K}
\end{array}\!\!
\right)
\end{equation}
[V] and [VI] model the term $\bm{\nabla}\cdot(h\bar{\bm{\sigma}})$ on the
left-hand side of (\ref{eq:FD_xmom_bal}) as $(1/2)\bm{\nabla}(\bar{\rho}
h^2 g_z)$ and neglect shear stresses in vertical planes (as virtually all
quasi-3D avalanche models do). The RKE-modified Voellmy friction law is
used to model the bed shear stress $\bss $---however, now with $R_V$
instead of $R_K$ determining the decrease of the friction parameters
[cf.\ Eqs.~(VI.35) and (VI.36)]. The slope-parallel diffusive flux of
RKE is neglected ($\bm{q} \approx 0$), as mentioned earlier. In
(\ref{eq:FD_FE_bal}), the dissipation $\varGamma$ is assumed proportional
to $R_K + R_V = R_K - M (k-k_0) g_z$ rather than to $R_K^{3/2}$ as
suggested by kinetic theory. Neither the different exponent nor the
appearance of $R_V$ in $\varGamma$ is incompatible with the general
framework because the latter does not specify the form of $\varGamma$,
but it is a clear departure from kinetic theory.

To the extent that Bartelt and co-workers assume the shear layer to be
infinitesimally thin, the supply of RKE could be described by setting the
boundary flux term $q_z|_0 = \alpha \bss  \cdot \ubb$. But it appears
more natural to regard $\alpha \bss \cdot \bm{u}|_0$ as the limit of
$h \overline{\bm{S}\cdot\partial_z \bm{u}}$ when the thickness of the
shear layer, $\delta_s$, tends to zero: In the shear layer, the shear
stress is approximately equal to $\bss $, and the shear rate is
$\partial_z \bm{u} \approx \ubb/\delta_s$. Integration over $z$ from 0
to $h$ then gives $\bss  \cdot (\ubb/\delta_s) \delta_s = \bss  \cdot
\ubb$. This would, however, impose $\alpha = 1$.

The extended RKE model of [V] and [VI] thus neglects all terms on the
second line of Eq.~(\ref{eq:FD_FE_bal}) except the third. The first and
fourth term describe RKE generation due to shear along vertical planes.
According to standard scaling arguments for shallow flows based on the
aspect ratio $\epsilon \ll 1$, $\ubb$ and $\partial_z$ are $O(1)$ while
$h$, $w$ and $\nablab$ are $O(\epsilon)$. Thus only the third term,
$h \overline{\bm{S}\cdot\partial_z \ubb}$, is $O(\epsilon)$ while the
others are $O(\epsilon^2)$ or $O(\epsilon^3)$ and would be negligible.
However, the second term, $h\overline{\sigma_{zz}\partial_z w}$, is
special in that it is present even if the flow does not deform in the
tangential directions of the incline, but changes density. Moreover, it
describes how RKE is transformed into potential (\ie, non-random) kinetic
energy as the flow expands in the bed-normal direction. This term thus
implements the feed-back mechanism governing density changes and must
not be discarded. A more detailed scale analysis of
Eq.~(\ref{eq:FD_FE_bal}) would need to introduce different time scales
and is beyond the scope of this paper, but will be invaluable in the
construction of an improved, consistent model.

Finally, comparing the last three equations in the system (VI.30),
(VI.37)--(VI.39),
\begin{align}
\partial_t h_\Phi + \nablab\cdot(h\bm{u}_\Phi)
&= w_\Phi ,                                     \label{eq:RKE_h} \\
\partial_t(M_\Phi w_\Phi) + \nablab\cdot(M_\Phi w_\Phi\bm{u}_\Phi)
&= N_K ,                                        \label{eq:RKE_w} \\
\partial_t N_K + \nablab\cdot(N_K \bm{u}_\Phi)
&= 2\gamma \dot{P} - 2N \frac{w_\Phi}{h_\Phi} , \label{eq:RKE_N}
\end{align}
with Eq.~(\ref{eq:FD_zmom_bal}) is not straightforward because the extended
RKE model here departs from the canonical approach based on the fundamental
balance equations. We first note that, if one assumes uniform density and
velocity profiles, one may insert Eq.~(\ref{eq:FD_mass_bal}) into
Eq.~(\ref{eq:FD_zmom_bal}) to obtain
\begin{equation} \label{eq:w_full}
D_t w = \frac{1}{h\rhob} \nablab\cdot(h\bar{\bm{S}})
- \frac{\sigma_{zz}|_0}{h\rhob} + g_z .
\end{equation}
Using the equation for $M_\Phi$ with $\dot{M}_{\Sigma\rightarrow\Phi}$,
which is identical to Eq.~(\ref{eq:FD_mass_bal}), the same procedure can
be applied to Eq.~(\ref{eq:RKE_w}) and yields (in our notation)
\begin{equation}
D_t w = - \frac{\sigma_{zz}|_0}{h\rhob} + f_z .
\end{equation}
The neglected term is $O(\epsilon)$, thus smaller than each of the
other two terms on the right-hand side of Eq.~(\ref{eq:w_full}), but
of the same order as $D_t w$ and the sum of the $O(1)$ terms. Equation
(\ref{eq:RKE_h}) appears to contain a misprint---$w_\Phi$ is the
centre-of-mass velocity, thus there should be a factor 2 on the
right-hand side. Even so, this equation is in conflict with the
kinematic boundary condition (\ref{eq:kin_BC}) because
$\nablab\cdot\bm{u}_\Phi \not\equiv 0$. In fact, tracing its
derivation in [VI], one sees that it \emph{should} be the kinematic
boundary condition rather than a dynamical equation. We have discussed
the reasons why Eq.~(\ref{eq:RKE_N}) is not valid in
Sec.~\ref{sec:variable_density}; comparing it with
Eq.~(\ref{eq:FD_zmom_bal}), it is apparent that it needs to be replaced
by a proper constitutive law for $\sigma_{zz}$ or, equivalently, the
pressure.


\end{document}